# Laser-driven few-cycle Terahertz sources with high average power

Robin Löscher[1], Tim Vogel[1], Samira Mansourzadeh[1], Mohsen Khalili[1], Alan Omar[1], Yicheng Wang[1], Martin Hoffmann[1], and Clara J. Saraceno[1,2*]

[1] Photonics and Ultrafast Laser Science, Ruhr-Universität Bochum, 44801 Bochum, Germany

[2] Research Center Chemical Science and Sustainability, University Alliance Ruhr, 44801 Bochum, Germany

(E-Mail: clara.saraceno@ruhr-uni-bochum.de)

**Ultrafast laser-driven terahertz sources are gaining in popularity in an increasingly wide range of scientific and technological applications. However, many fields continue to be severely limited by the typically low average power of these sources, which restricts speed, signal-to-noise ratio, and dynamic range in numerous measurements. Conversely, the past two decades have seen spectacular progress in high average power ultrafast laser technology based on Ytterbium lasers, rendering hundreds of watts to kilowatts of average power available to this community to drive THz sources. This has opened the young field of high-average-power laser-driven THz time-domain spectroscopy, which holds the potential to revolutionize the applications of THz time-domain systems. In this perspective article, we discuss this young field and emphasize recent advancements in broadband terahertz sources utilizing high-power Yb-based ultrafast lasers as drivers, which are nearing watt-level average power. We discuss various approaches explored thus far, current challenges, prospects for scaling, and future research areas that will accelerate their implementation in applications.**

## I. Introduction

Coherent sources producing ultrashort, few-cycle pulses of terahertz (THz) radiation have become well-established in an immense variety of scientific fields, spanning physics, chemistry, material science, engineering, biology, and medicine[1]. They allow for steering and probing the dynamics of a variety of low-energy phenomena in condensed matter, ranging from ultrafast dynamics of charge carriers[2,3], phonons and spins[4,5], conductivity[6,7], to intermolecular dynamics of liquids in biological systems[8,9]. The dynamics accessed by these short THz pulses can even nowadays be accessed with sub-wavelength resolution using near-field techniques[10], now achieving angstrom spatial resolution[11–13]. Furthermore, ultrafast laser-driven THz sources are powerful tools for material identification and imaging[14,15], and are starting to be increasingly deployed in industrial inspection scenarios, in particular for quality control of layered materials[16]. Given the immense variety of applications and fields that these sources serve, it is natural that advances in their performance are intimately tied to advances in THz science. Among the multitude of technologies advancing the THz spectral region in general[1], we focus our attention here on ultrashort, broadband THz pulses, using femtosecond laser pulses as drivers. Such short THz pulses with high powers can also be reached with superradiant accelerator-based sources[17], which also serve this community. However, we focus here on recent advances in laser-driven, tabletop sources that could be widely adopted in individual labs, as these have seen most pronounced and fast-paced progress in last years.

The most commonly used techniques (Table 1) to generate ultra-broadband THz pulses using ultrafast lasers are down conversion in materials with $\chi^{(2)}$ nonlinearity using intra-pulse difference frequency generation (optical rectification)[18–20], photocurrents in biased semiconductors and ionized gaseous media[21,22], and more recently, spintronic THz emitters[23–26]. Each one of these methods has distinct advantages and disadvantages and gives access to different parameters in terms of bandwidth, pulse duration, electric field strength, beam quality and polarization states; the choice of one or the other depends on the desired parameters for a given application and are strongly constrained by available laser systems and parameters at hand for driving the conversion mechanism. In this context, progress in laser technology is immensely relevant to support advances in ultrafast THz technology. While all the techniques mentioned above have seen continuous progress, effectively closing the now outdated term "THz gap," a significant technical challenge remained for many years: the typical low average power of the sources.

In the last few years, however, the rapid advancement of high-power ultrafast laser technologies based on ytterbium (Yb)-doped solid-state lasers has enabled us to make significant progress in overcoming this hurdle. Ultrafast laser systems with tens to hundreds of watts of average power are now commercially available with industrial-grade stability, and pulse compression methods are transforming these lasers into the new workhorse of broadband THz science. In this new context, this perspective article summarizes recent findings on applying these laser systems to generate high-average-power broadband THz radiation and provides a forward-looking perspective on the current challenges and next steps ahead. Rather than focusing on one specific application and its needs, we cover the common need for high average power across the different source types and discuss the various challenges and differences that arise in the generation mechanisms when increasing the driving average power.

Table 1: Average power scaling attempts reported in literature for each specific technology, all based on ytterbium drivers.

| Method | Schematic illustration in the case of Yb drivers | Peak field [kVcm$^{-1}$] | Average power [mW] | Repetition rate of corresponding reference (left) [kHz] | Energy [J] | Bandwidth* [THz] | Optical-to-THz efficiency | Typical driving pulse energy for efficient conversion [J] |
|---|---|---|---|---|---|---|---|---|
| Spintronic emitters | 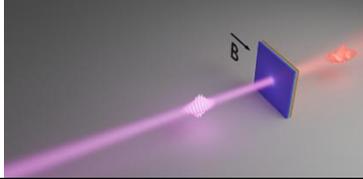 | 10, Ref.[27] | 0.059, Ref.[28] | 400 | 97.5·10$^{-12}$, Ref.[28] | 10, Ref.[27] | 5.6·10$^{-6}$, Ref.[28] | 10$^{-8}$ … 10$^{-3}$ |
| Semiconductor or photoconductive emitters | 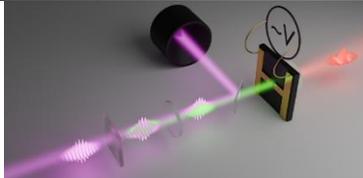 | 31, Ref.[29] | 6.7, Ref.[29] | 400 | 1,7·10$^{-9}$, Ref.[30] | 4, Ref.[31] | 3.8·10$^{-4}$, Ref.[29] | 10$^{-9}$ … 10$^{-6}$ |
| Non-collinear optical rectification i.e. tilted-pulse front | 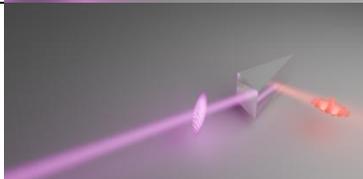 | 150, Ref.[32] | 643, Ref.[33] | 40 | 436·10$^{-6}$, Ref.[34] | 4, Ref.[35] | 3.8·10$^{-2}$, Ref.[36] | 10$^{-5}$ … 10$^{-1}$ |
| Collinear optical rectification i.e. organic crystals | 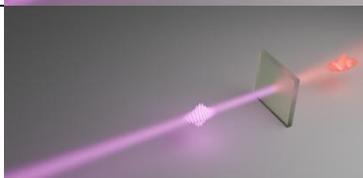 | 200, Ref.[37] | 11, Ref.[38] | 100 | 3.6·10$^{-6}$, Ref.[39] | 19, Ref.[39] | 5.5·10$^{-3}$, Ref.[39] | 10$^{-7}$ … 10$^{-3}$ |
| Two-color plasma | 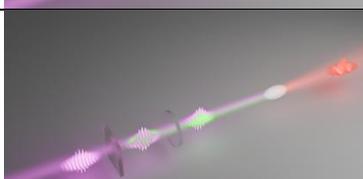 | 165, Ref.[40] | 640, Ref.[41] | 500 | 1.3·10$^{-6}$, Ref.[41] | 60, Ref.[42] | 1.1·10$^{-3}$, Ref.[41] | 10$^{-4}$ … 10$^{-1}$ |

*Typically reported bandwidths using different detection methods. Examples reflect best reported values obtained for each technology, not necessarily those measured at high average power.

## II. Relevance of increasing the average power in THz time-domain measurements

To grasp the immense relevance of increasing the average power of laser-driven THz sources, it is essential to understand both the detection schemes used to measure temporal transients and the fundamental relationship between signal quality and measurement speed. A typical THz transient generated by an ultrafast laser exhibits few-cycle pulse characteristics with extremely wide bandwidths spanning sometimes multiple octaves. The THz pulse train is phase-stable as all generation mechanisms follow closely the pulse envelope of the driving laser, which offers high stability. This represents one of the primary advantages of these sources, as it enables coherent detection schemes such as electro-optic sampling (EOS)[43], with much higher sensitivity than the commonly used Fourier-transform infrared spectroscopy (FTIR) technique, and with the additional possibility of extracting not only the amplitude, but also the phase of the THz electric field transient. While single-shot detection techniques exist and are currently the topic of intense exploration[44–47], the predominant and most performant approach in terms of dynamic range (DR) employs multiple time-delayed pulses from the driving laser to fully reconstruct the THz electric field in both amplitude and phase with high fidelity (see Fig. 1(a)). This combination of emitter and coherent detection forms the foundation of Terahertz time-domain spectroscopy (THz-TDS).

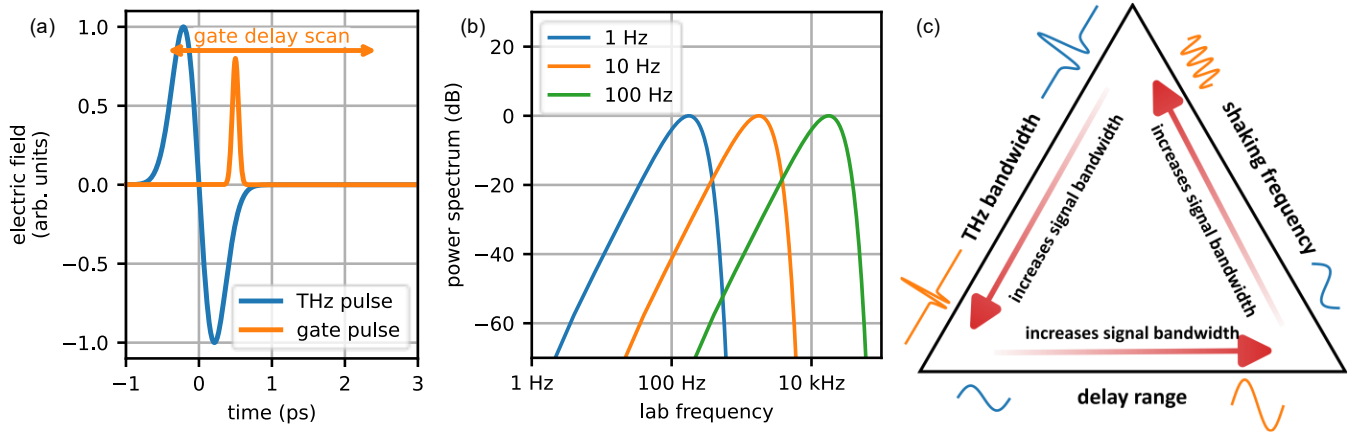

Figure 1: (a) Single-cycle THz and gate pulses illustrating the delay scanning in conventional TDS. (b) Influence of the shaker frequencies (legend) on the signal bandwidth of a simulated THz pulse with 4 THz bandwidth. (c) Relation between THz bandwidth, delay range, and shaking frequency for sinusoidal delay lines. (b) and (c) adapted from Ref.[48].

Understanding the signal-to-noise ratio (SNR) becomes crucial when evaluating the performance of high-average-power systems. The THz community employs DR to describe what other scientific communities term SNR[49]. State-of-the-art THz-TDS setups based on this coherent detection principle reach a peak dynamic range in the frequency domain of >>100 dB[50–52]. While increasing THz pulse energy enhances signal strength for constant noise levels, pursuing THz average power rather than pulse energy alone requires to consider multiple aspects regarding measurement speed and performance, guided by the given needs of an application.

For equivalent pulse energies, increasing the repetition rate provides multiple advantages that directly impact experimental capabilities. Higher repetition rates enable the measurement of faster processes, particularly in single-shot experiments where samples undergo irreversible changes, preventing measurements where repeatable signals are essential such as the pump-probe scheme[53]. The Nyquist criterion fundamentally restricts the signal bandwidth to values below half the laser repetition rate. Elevated repetition rates enable a broader signal bandwidth within the laboratory time frame, allowing for more efficient conversion from light time to lab time during the detection process. For example, if one considers a mechanical delay stage operating with continuous movement profiles covering approximately 50 ps of delay while sampling a few-THz broadband source using a conventional 1 kHz titanium (Ti):Sapphire laser system, the Nyquist-imposed bandwidth limitation restricts scan rates to approximately 1 Hz, yielding a maximum of two THz traces per second during forward and backward movements (see Fig. 1(b)). Commercial delay lines can achieve scan rates of 20 Hz or higher, and increased laser repetition rates enable proportionally faster sampling until mechanical limitations become the primary constraint, as illustrated in Fig. 1(c). Equivalent-time sampling techniques using two repetition rate shifted lasers (often referred to as asynchronous optical sampling in the THz field[54]) represent a further improvement for fully exploiting high-repetition-rate lasers, operating near the Nyquist limit even at elevated repetition rates. Recent implementations in single-cavity geometries represent an emerging technology with average power capabilities still under development[55].

High repetition rate operation provides additional noise reduction beyond simple averaging. Elevated repetition rates distribute signal content across broader laboratory frequency ranges. Since laser systems and electronic components exhibit increased 1/f noise at lower laboratory frequencies, expanding the relevant signal bandwidth to higher frequencies results in cleaner trace acquisition. This frequency domain advantage complements the temporal benefits of accelerated scanning.

Acquiring numerous traces per second through high repetition rate systems enables advanced statistical methods and correction algorithms[48,56,57]. Large trace ensembles facilitate sophisticated post-processing techniques that enhance measurement quality and extract additional information from acquired data.

THz average power represents the direct multiplication of pulse energy and repetition rate, fundamentally determining the signal strength achievable within short measurement durations. However, very high power systems can introduce complications through increased beam pointing instabilities, amplitude fluctuations, and other effects that elevate noise floors[58], requiring careful design to maintain anticipated SNR improvements.

## III. High-power laser technology advancements to support THz science

For the past decades, two families of ultrafast lasers have dominated the generation of broadband THz pulses: on the one hand, applications demanding strong THz fields relied on Ti:sapphire-based chirped-pulse amplifiers (CPA)[59] and their high driving pulse energies - typically up to several mJ and short pulse durations well below 30 fs at the table top; on the other hand, for applications in linear spectroscopy, Ti:sapphire oscillators with high repetition rate (typ. 80 MHz) and very short pulses durations <10 fs were commonly deployed in the scientific community. Meanwhile, commercial and robust low-noise, low-cost Erbium (Er) fiber lasers are used in most commercial broadband THz spectrometers, both for industrial applications such as inspecting paint layers or cultural artwork[60,61], but also in scientific applications[62]. However, both Ti:sapphire lasers and Erbium fiber lasers are challenging to scale in average power, which partly explains the lack of progress in the generation of high-power THz.

Regarding Er-doped fiber lasers, research has focused on low-noise and low-cost approaches that enhance the DR at low average powers. Power scaling Er-doped lasers is challenging, due to a substantial quantum defect (ratio of pump wavelength to laser wavelength) in the laser gain electronic laser levels; however, developments using Er:Yb co-doped fiber systems demonstrate scalability and could support further advances[63].

On the other hand, Ti:sapphire has extremely favorable properties as a laser gain material for ultrafast operation; in particular, an extremely broad gain bandwidth, enabling the generation of few-cycle pulses from oscillators and very broadband operation (>30 nm) from high-gain amplifiers, i.e. regenerative amplifiers. However, whereas the gain material itself has excellent thermal conductivity (33 Wm$^{-1}$K$^{-1}$)[64], it exhibits a relatively short gain lifetime, resulting in a high saturation power. High doping levels are not possible while maintaining good crystal quality; therefore, longitudinally pumped bulk crystalline media in combination with small pump spots are required to achieve efficient laser operation. The poor heat extraction in this geometry, combined with a large quantum defect, strongly limits the average power achievable from these systems to a few watts at room temperature in optimum cases. Cryogenically cooled systems can reach higher powers of tens to hundreds of watts[65] but with very complex cooling designs, primarily aiming at facility-type infrastructures with ultra-high peak powers.

## III.a Ytterbium-doped gain media: the new high-power ultrafast workhorse

Laser gain materials doped with Yb operating at a central wavelength around 1 µm (typical central wavelength of 1030 to 1050 nm) offer very high quantum efficiency (> 90%), alongside with low upconversion and excited-state absorption in most commonly used hosts, for example Yb:YAG[66]. The typically high gain cross-section and long lifetimes of Yb-doped gain materials allow for cost-efficient direct pumping by low-brightness laser diodes, reaching optical-to-optical efficiencies exceeding 80%[67], reducing the heat load by the pump significantly when compared to Ti:sapphire[68]. These advantages make Yb lasers ideally suited to combine with geometries that offer good heat extraction via an improved surface-to-volume ratio of the gain medium, for example, in slab, fiber, and thin-disk geometries. This has allowed for immense progress in the performance of continuous-wave (cw) lasers[69] but also systems with ultrashort pulses: average powers exceeding kW-level in sub-ps operation are now routinely obtained at repetition rates from the kHz to multi-MHz[70–74]. A large majority of these advances were based on a chirped-pulse amplification scheme to achieve high pulse energies. For instance, sub-ps pulses with up to 720 mJ of pulse energy at 1 kHz using thin-disk amplifiers[75], and up to 10.4 kW of average power, 130 µJ pulse energy, 254 fs pulse width from a fiber-based architecture[74] were demonstrated. Using the thin-disk architecture, high-power, mode-locked oscillators have reached the 100 µJ level with average powers up to 550 W at 5 MHz repetition rate[76]. It is noteworthy that some of these kW-class ultrafast lasers were demonstrated 15 years ago[70]. This has led to the availability of industrial-grade, high-power Yb ultrafast lasers with hundreds of watts of average power, making them accessible to the scientific community, including the THz community. However, their widespread adoption in scientific applications, which very commonly require the generation of secondary radiation such as THz light, has progressed relatively slowly in comparison. One reason for this is the typically narrow spectral gain bandwidth of Yb-doped gain media, which makes it challenging to obtain short pulse durations directly from high-power laser systems. Typical commercial laser systems exceeding 100 W of average power achieve pulse durations in the range of 200 fs to ps, as illustrated in Fig. 2. In most laser-driven THz generation methods, external pulse compression techniques are therefore essential: on the one hand, short pulse durations and their broad bandwidths are required to generate and detect correspondingly wide bandwidths in the THz domain; on the other hand, for many generation methods – for example using nonlinear crystals or gas plasmas – the driving laser peak power is a critical factor for reaching good conversion efficiency. Reaching these high peak powers becomes increasingly challenging at very high repetition rates (i.e. MHz and beyond) where pulse energies are more moderate even at high average power, making these pulse compression methods even more critical for this repetition rate regime. Along with progress in the laser systems themselves, pulse compression methods compatible with high average power have also significantly progressed over the past few years, and have correspondingly accelerated the adoption of Yb-doped lasers for THz generation. As a key enabling development, we detail some of the latest advancements in this area below.

## III.b Pulse compression methods to support the application of Yb lasers for THz generation

At the wavelength of Yb lasers, the most commonly used pulse compression method nowadays is spectral broadening by self-phase modulation (SPM), followed by dispersive compensation and corresponding temporal compression. Another commonly used technique to achieve pulse shortening is optically pumped chirped pulse amplification (OPCPA), nowadays increasingly pumped by high-power Yb systems. These systems start to reach very high average powers as well[77], but they are significantly less efficient, with typical conversion efficiencies from pump to compressed pulses in the order of the percent level, and they are significantly more complex to implement. These systems mostly excel in applications requiring few-cycle durations and excellent pulse contrast, such as attosecond science. Whereas this performance certainly remains relevant for certain THz generation schemes, for example for symmetry breaking without two-color fields in a plasma, the most relevant compression method for advancing THz science at present are those based on SPM spectral broadening. In particular, schemes with low loss, large compression factors for long input pulses, good beam quality and robustness regarding thermal loads for average power scaling are more widely needed. Other recent reviews have focused on detailed technical aspects and progress in pulse compression methods[78], we therefore here only present a concise overview of available technologies that have supported and will continue to support advances in different THz generation techniques.

Methods for spectral broadening can be categorized by the peak power range of optimal operation. For many decades, waveguide-based methods (solid-core or hollow-core fibers) dominated nonlinear post-compression. However, solid-core fibers are limited by the critical peak power for self-focusing to a few MW in silica, making this solution viable only at very low pulse energies. High average power laser systems with tens to hundreds of watts easily reach µJ-level pulse energies and peak powers exceeding tens of MW, even at multi-MHz repetition rates; therefore, solid-core fibers quickly proved inadequate for the fast-paced progress of the sources. The use of gas-filled waveguides with lower nonlinearity, such as gas-filled hollow-core photonic crystal fibers (HC-PCF)[79,80] was therefore explored early on. They provide large and spatially homogeneous spectral broadening; however, they are challenging to scale in average power. Nearly perfect coupling efficiency is required, imposing nearly ideal diffraction-limited beam quality, which is often not fulfilled in very high average power lasers or amplifiers. Furthermore, maintaining coupling into these waveguides and avoiding damage at extremely high average powers for long-term experiments represents a major engineering challenge.

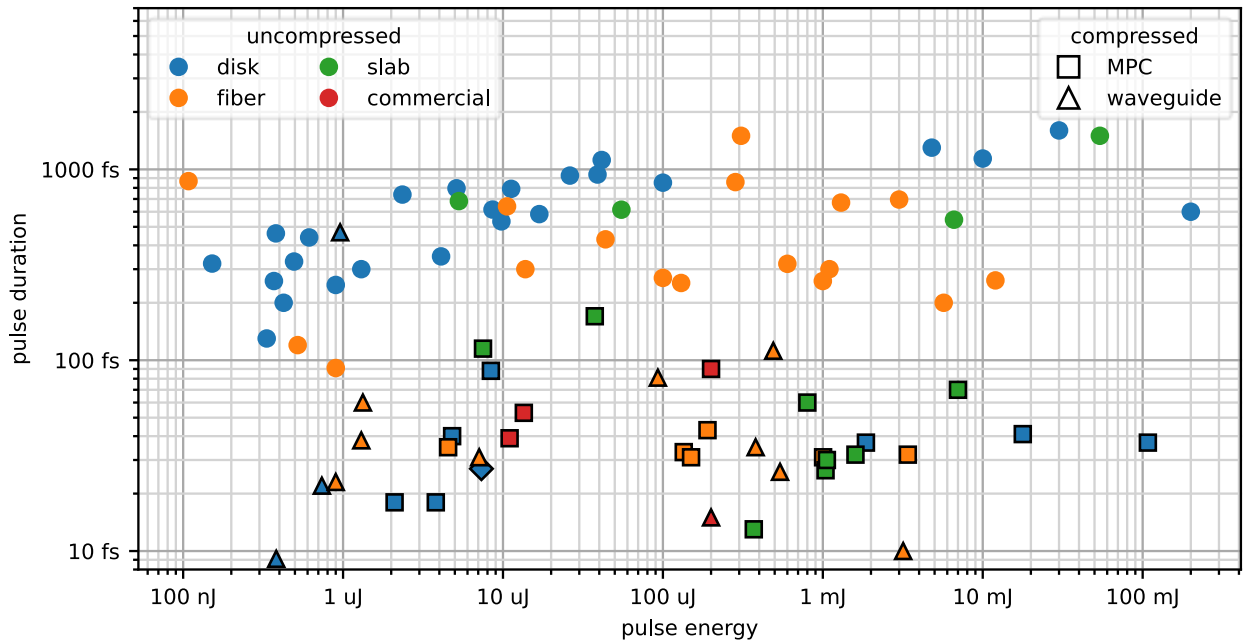

Figure 2: Overview of Yb-based laser systems and post-compression techniques operating at average powers exceeding 10 W. Post-compression methods enable sub-100 fs pulse generation across average powers ranging from tens of watts to the kilowatt level, with corresponding pulse energies spanning microjoules to multi-millijoules. These capabilities address diverse application requirements for terahertz generation and detection, as detailed in Section IV.

For higher pulse energies in the mJ-class, hollow-core fiber (HCF) compression, utilizing capillaries, is a commonly used method, enabling pulse durations down to two-cycle duration with good beam quality[81–83], and good average power handling, so far tested up to 300 W[84]. The main drawback of capillaries is their extreme bending sensitivity with respect to loss, requiring very large footprints and engineering efforts to handle average power. Nevertheless, this technology continues to progress and is actively being deployed in the context of THz generation [39,42].

More recently, the use of free-space alternatives has gained in popularity for high-average-power lasers. In particular, the use of a Herriott-type multi-pass cell (MPC) for spectral broadening was an important breakthrough to bring Yb systems to THz generation applications, and the large majority of current results are using this method[38,40,85–88]. This technique uses repetitive propagation through a nonlinear medium (which can be bulk, gas, or a combination of both, depending on the pulse energy), achieving spatially homogeneous spectral broadening[86,89] when designed carefully using multi-dimensional pulse propagation code[90]. It is particularly well suited for high compression ratios and rather long input pulses and offers the practical advantage of free-space propagation, making it compatible for high average power. The first demonstration of this technique dates back to 2016[91] and dominates research in compression techniques for high average power lasers (see Fig 2). The scaling laws and advantages of MPC based compressors have been discussed extensively in the last years[78,86] and their exploration constitute a currently very active research field[92,93] beyond spectral broadening only. They represent an extremely versatile platform that supports kW levels[94,95] and pulse energies from µJ to multi-hundred mJ[95,96]. Multistage MPCs can even now reach few-cycle durations, approaching those of capillaries and OPCPA with main limitations being the bandwidth of the dielectric mirrors typically preferred for high average power applications.

Finally, multi-plate compression in solid media is also an alternative, where the nonlinear spectral broadening happens in consecutive thin plates[97,98]. In this method each subsequent plate induces moderate nonlinear broadening via SPM, and propagation in between allows to circumvent beam degradation. This can be considered an extension of the multi-pass cell concept, better suited for very short input pulses and higher energies, where each nonlinear step needs separate adjustment[98]. This method is mostly used as a final compression stage with short input pulses to reach few-cycle pulse durations[96,99] and has therefore rarely[100] been deployed for THz generation so far.

## IV. State of the art of THz generation methods at high average power

These breakthroughs in Yb-based ultrafast laser technology have accelerated the realization of broadband THz sources at very high average power in several schemes over the past few years, now approaching the Watt level in record-holding systems. An overview of the current state of the art is presented in Fig. 3, illustrating that these advances are occurring at the crossing between the traditional Ti:sapphire CPA based strong-field THz sources—usually for applications as pumps in spectroscopy and fundamental science—and high repetition rate setups driven by Ti:sapphire oscillators and Er-doped fiber lasers – mostly for application in linear spectroscopy. It becomes evident that Yb technology is playing a key role in merging these areas: setups requiring THz individual pulse strengths can now operate at high repetition rate and benefit from faster acquisition, while setups in linear spectroscopy can also leverage stronger single-pulse strengths to enhance SNR or consider pumps at very high repetition rates. This has made Yb technology connect disciplines and know-how that were often disconnected.

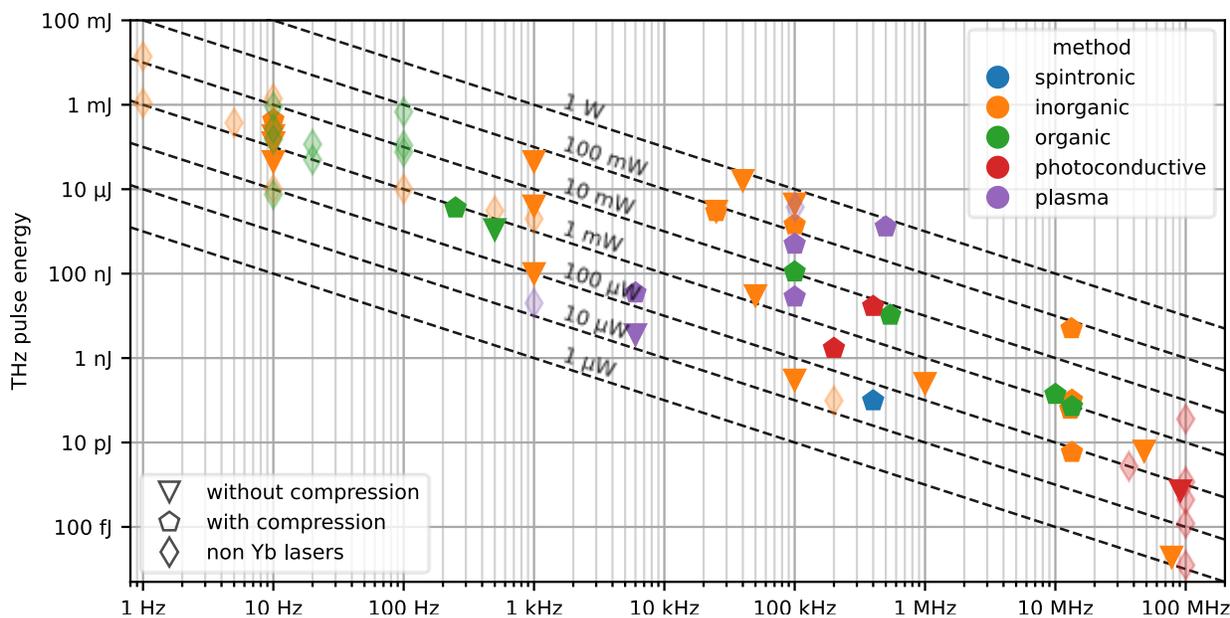

Figure 3: Overview graph for THz sources based on ytterbium with compression[28,30,32,34,35,38–42,79–87], and without compression[29,31,33,34,36,37,88–96]. Results not achieved with Yb-based drivers in can be found at Ref.[32,132,139,178,247–267,267–271].

We detail below the different THz generation methods that have been studied in combination with high-power Yb drivers and the main findings so far. We organize the presentation starting with those generation techniques where most research has happened in recent years and thus are becoming increasingly mature for application, starting with optical rectification, to other techniques that present more challenges, where progress is most recent, and more research can be expected in the coming years. We particularly focus on different findings related to average power-related changes in the generation mechanisms themselves, rather than present general aspects that have been extensively discussed in the literature.

## IV.a Optical rectification

Optical rectification (also referred to as intra-pulse difference frequency generation) in $\chi^{(2)}$ materials is perhaps the most used technique for generating THz pulses, particularly in the scientific community, to reach wide bandwidths and/or strong fields for spectroscopy. The general aspects and scaling laws for optical rectification have been extensively discussed in the literature – we therefore do not cover them here and refer the reader to previous work[101] and instead focus on specific aspects of the THz pulse generation that are affected by operating at high average power. In this case, two material properties become particularly important: low pump absorption and good thermal conductivity, which enable the efficient dissipation of absorbed energy and prevent thermal accumulation and thermally induced damage (Table 2).

Other critical aspects that need to be considered are temperature-dependent material properties, thermal lensing, and depolarization. These thermal effects are challenging to tackle in both inorganic and organic nonlinear crystals, and their study is gaining in relevance. We discuss here specific findings made so far in various nonlinear crystals for optical rectification, and include new measurements performed in one of the most prominent materials for THz generation: lithium niobate (LiNbO$_3$, LN).

Table 2: Comparison of key material properties for crystals used in terahertz generation by optical rectification at 1030 nm pump wavelength.

| Crystal | Thermal conductivity [Wm$^{-1}$K$^{-1}$] | Bandgap at 300 K [eV] | α at 1 THz [cm$^{-1}$] | $d_{eff}$ at ~1 μm wavelength [pmV$^{-1}$] |
|---|---|---|---|---|
| GaP | ~500 at 90 K[102] | 2.26[103] | 0.2[104] | 24[105] |
| BNA* | 0.18 at RT[106] | 3.79[107] | 4[108] | 234[109] |
| LN | ~100 at 80 K[110] | 3.95[111] | 17[104] | 160[105] |

*BNA has parameters representative for the organic crystal family

## IV.a.i Inorganic crystals
**Gallium phosphide**
One of the main critical criteria for efficient THz generation using optical rectification is broadband group-to-phase velocity matching between the near-infrared pump and the THz pulses[101] for a significant crystal length. In this regard, gallium phosphide (GaP) plays a vital role for Yb drivers, as it offers broadband collinear velocity matching for THz frequencies at driving wavelengths of ~1 μm. Furthermore, large samples of high quality can be manufactured and are widely commercially available. The material combines several additional favorable properties, including a relatively high second-order nonlinear coefficient, broad transparency from ~0.55 μm to ~1.6 μm, and relatively low terahertz absorption coefficients at room temperature[112]. With this in mind, GaP has been explored with Yb systems early on[112,113] and power scaling with >100 W of driving average power at high repetition rates (13 MHz) has been demonstrated[114]. In this high average power regime, limitations were explored in detail, and it was shown that nonlinear absorption remains the limiting factor for scaling, severely restricting the intensities applicable on the crystal and thus the conversion efficiency[115]. In fact, the main drawback of GaP for pumping in the near infrared is its small bandgap (2.26 eV), making multi-photon absorption the main limiting factor for scaling. With high power Yb drivers, three-photon absorption at these pump wavelengths create both a strong thermal load and free carrier generation, which further increases THz absorption. It has also been shown that cryogenic cooling only partially relaxes thermal effects yielding a >30% enhancement in terahertz emission and extended spectral coverage at 80 K[115]. However, efficiencies remain very low in all cases, limiting the THz power to the mW level with very high input powers >100 W. One advantage, however, remains the wide bandwidths achieved, typically up to 7 THz before being limited by phonon absorption[116]. Nevertheless, albeit at reduced conversion efficiency, GaP is a very robust material, sustaining high powers and intensities without damage at elevated temperatures, still making it an attractive material for many applications in practice.

**Lithium Niobate (LN)**

Compared to GaP, lithium niobate offers several advantages: a wide bandgap energy[111] (indirect: 3.95 eV, direct: 4.12 eV) allowing for significantly higher intensities before multi-photon absorption becomes limiting, as well as a high $\chi^{(2)}$ coefficient. However, group-to-phase velocity matching is not fulfilled collinearly for significant crystal thicknesses; therefore, alternative strategies are required. The most popular technique when high driving pulse energies are available is the tilted-pulse front geometry[117]. Other geometries, such as the Cherenkov scheme[118–120] in a slab geometry, have been implemented to circumvent this problem in the case of reduced input pulse energies[121], but have so far not been explored with high average power from Yb-based lasers.

Lithium niobate tilted-pulse front (LN-TPF) based THz sources have seen most progress in achieving high average powers driven by Yb-based laser systems. Most progress in this area has been achieved using mJ-class CPA laser systems operating between 10 kHz and a few hundred kHz, offering similar pulse energies to Ti:sapphire amplifiers at the mJ level, but at one to two orders of magnitude higher repetition rates. In 2020, Kramer et al. presented a first result reaching 144 mW of average power driven by a Yb:YAG slab amplifier[122] with 375 W of pump power at 100 kHz, with a room-temperature LN crystal. More recently, our group extended these studies, reaching up to 643 mW of THz average power with an Yb-based thin-disk regenerative amplifier at 40 kHz, and 450 mW at 100 kHz[33], using cryogenic cooling at 80 K. Other results using more moderate driving average powers of tens of W are also increasingly being reported in this repetition rate region, now more routinely enabling tens of mW of average power to become more widely available[35,123].

Concurrently, the MHz repetition rate region was explored in this scheme, employing a pump average power of >100 W[124]. This regime has different challenges to the above: since much lower input pulse energies are available to drive the conversion mechanism, small laser spot sizes are required, which poses additional constraints for LN-TPF to avoid walk-off in the crystal[125]. In this regime, 66 mW of average power was demonstrated at 13 MHz repetition rate, which nevertheless remains the highest power THz source so far achieved at >MHz repetition rate[124].

Forward-looking developments

While these results suggest that W-class THz systems and beyond are attainable, there are many difficulties ahead, and many scaling laws remain to be understood.

In all excitation regimes where laser systems with an average power of more than 100 W were employed, thermal effects were evidently a limitation to conversion efficiency. Experiments performed at high repetition rates at a given pulse peak power systematically achieve lower conversion efficiencies in optimized conditions. The exact limiting mechanisms are very complex to disentangle and are strongly dependent on both the driving pulses and the geometry and imaging aspects in the LN-TPF. Here, we present some recent findings, aiming to better understand thermal effects and other high-average-power-related limitations in this scheme and further discuss future steps to continue this scaling trend.

In recent experiments, we performed measurements to quantify thermal effects in these crystals at high excitation average power in the regime of high pulse energies (>mJ), at repetition rates of 10 to 100 kHz. The experimental setup is a modified version of the setup described in Ref.[33] and is schematically illustrated in Fig. 4(a). From previous experiments[33], using crystals with an aperture of 20 × 20 mm$^2$, we were limited by the available aperture of the crystal to scale pulse energy to the available 50 mJ. We therefore recently acquired larger stoichiometric LN crystals with an input aperture of 30 × 30 mm² (OXIDE Corporation) and increased the crystal and grating apertures to reduce the applied fluence. The spot diameter of the pump beam at the LN crystal was ~5 mm and ~10 mm at the $1/e^2$ level for the x and y axes, respectively, corresponding to a maximum peak intensity of about 29 and 163 GW/cm² at 100 and 10 kHz of repetition rate, respectively. The crystal is placed in a cryogenic environment utilizing a heater module with a liquid-nitrogen flow to regulate the temperature of the upper copper plate ($T_1$, Fig. 4(b) and 4(c)). The maximum cooling power of ~7 W limits the applicable pump power in this study. A secondary temperature sensor monitors the lower plate ($T_2$). We investigate two mount configurations (Fig. 4(c)): asymmetric cooling, where heat transfer occurs through the crystal and spring-loaded screws between plates, and symmetric cooling, which adds a copper cold bridge at the unused edge of the crystal.

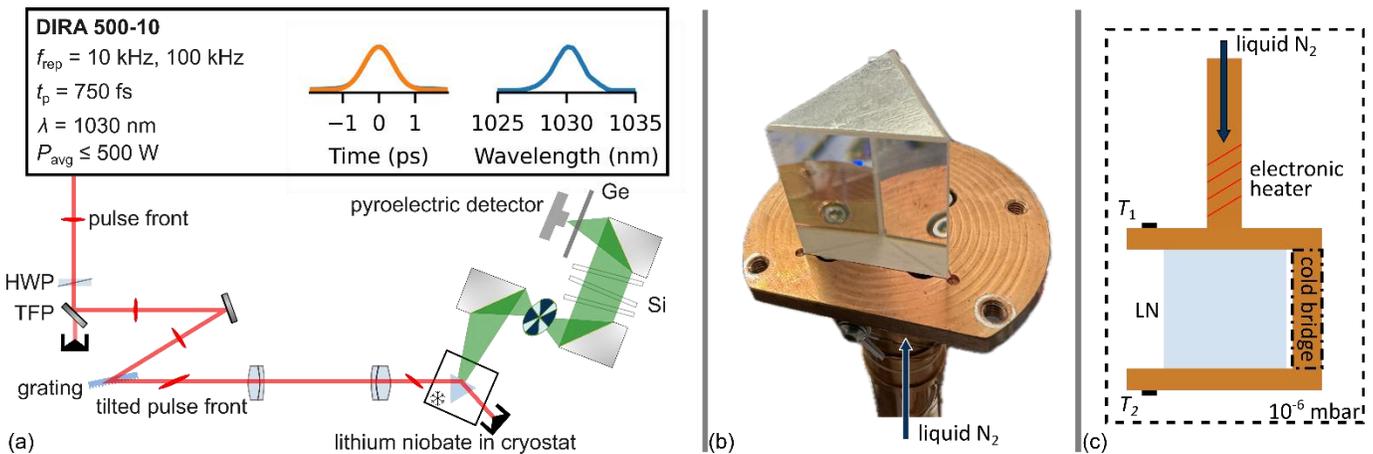

*Figure 4: (a) Experimental setup of the tilted pulse-front THz source with a large aperture LN crystal for high average powers. HWP: half-waveplate, TFP: thin-film polarizer, Si: silicon wafer stack, Ge: germanium wafer. (b) Picture of the LN crystal. (c) Schematic of the crystal mount design with temperature sensors $T_1$ and $T_2$. The copper cold bridge for the symmetric cooling configuration is shown as the dotted-dashed region.*

collimated beam path[126]. A germanium wafer positioned before the detector filters residual pump radiation and higher harmonics. All power measurements are corrected for filter attenuation.

Figure 5 shows strong relaxation times in the THz power using the asymmetric cooling configuration. The THz output power decays exponentially to a thermal equilibrium over hundreds of seconds, with 30% to 50% power reduction from initial to steady-state conditions. At the same average power, the 10-kHz case results in significantly more pronounced thermal effects, indicating that heating originates in a large part from nonlinear absorption. On the other hand, when comparing the 10 kHz case and 100 kHz case at constant pulse energy, the 100 kHz case introduces remarkably more pronounced thermal effects, indicating cumulative heating in the crystal. The exact causes of heating in all these regimes

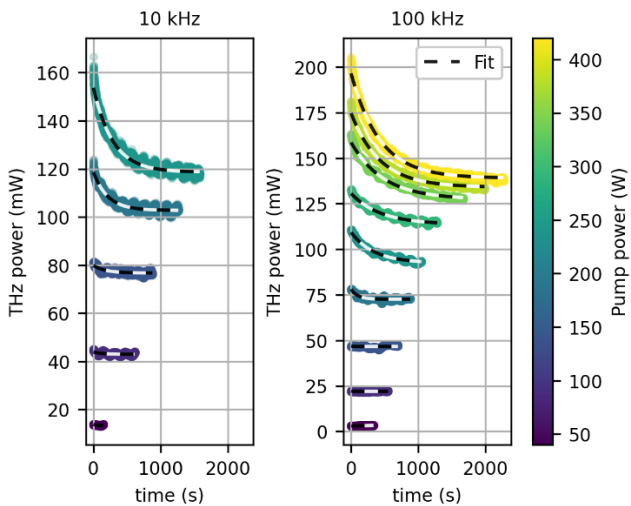

Figure 5: Temporal evolution of THz power showing thermal rolloff effects in LN crystals at 10 kHz (left) and 100 kHz (right) repetition rates. Exponential decay to thermal equilibrium occurs over hundreds of seconds, with 10 kHz showing more severe rolloff than 100 kHz, indicating peak intensity-dependent thermal effects. Black-white dashed lines show exponential fits to the data.

We employ a Yb-based thin-disk regenerative amplifier operating at 1030 nm with ~750-fs pulse duration and 500-W maximum average power. This provides pulse energies of 50 and 5 mJ at repetition rates of 10 and 100 kHz, respectively. The average power is controlled using a half-waveplate and polarizer, calibrated at the crystal position. The THz radiation is imaged in a 6f configuration and focused onto a pyroelectric detector (SLT GmbH, THz20), calibrated by the German national metrology institute (PTB) for a flat spectral response within our source bandwidth. Signal modulation is achieved using a chopper at the first THz focal spot. To prevent detector saturation, we attenuate the THz power with Si wafers in the remain to be studied in detail, possibly in a wider context using LN thin slabs and including detailed thermal simulations. Temperature gradient measurements across the crystal (Figure 6 and 7) reveal substantial thermal gradients of 40 to 60 K between the actively cooled and passively cooled surfaces, with final temperatures occurring on the same time scales as the THz power drop, clearly indicating a correlation between the two.

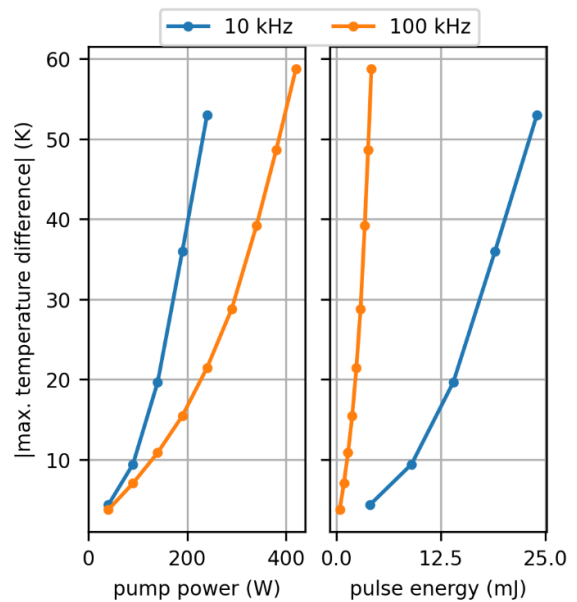

Figure 6: Crystal temperature gradients between actively cooled and passively cooled surfaces as functions of pump power (left) and pulse energy (right) for 10 kHz and 100 kHz of repetition rate. Temperature differences reach tens of K at moderate powers, with steeper scaling at 100 kHz demonstrating the critical role of thermal recovery time between pulses in high-repetition-rate operation.

A higher crystal temperature can have several detrimental effects on THz power: on the one hand THz absorption increases with temperature at these frequencies, however, the observed magnitude of the THz power drop of up to 70 mW (at 100 kHz, 420 W of pump power) cannot only be explained by this effect alone. On the other hand, we also observed experimentally that the optimal grating angle for tilted pulse front phase matching shifts significantly with crystal temperature, showing that thermal effects indirectly reduce conversion efficiency by altering the refractive index. This is shown in Fig. 8, where we heat the crystal using the electric heater in the cryostat

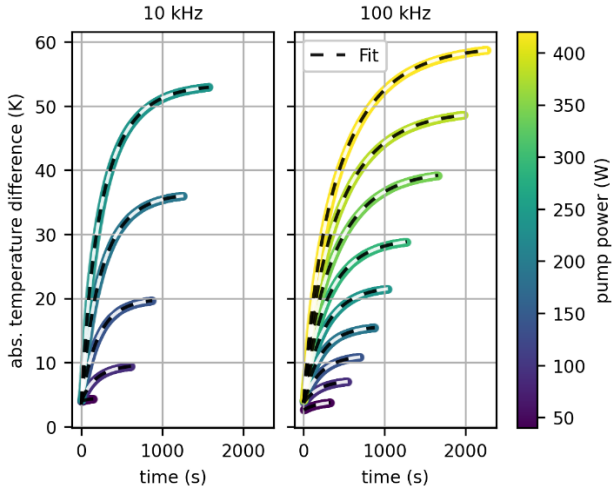

Figure 7: Temporal evolution of the absolute temperature difference $|T_2 - T_1|$. Double-logistic growth to thermal equilibrium occurs over hundreds of seconds, with 10 kHz showing more severe heating than 100 kHz, indicating peak intensity-dependent thermal effects. Black-white dashed lines show double-logistic fits to the data.

and observe the variation of the optimal grating angle.

When improving the cooling geometry with the cold bridge (Fig. 4(c)), thermal relaxation at 100 kHz was drastically reduced. This improvement resulted in the complete elimination of the power-dependent grating angle optimum (Fig. 9). This is a very promising finding that suggests further average power scaling to the W-level, i.e., kW-level excitation at MHz repetition rates may be feasible with current setups.

The 10 kHz case with higher pulse energy is considerably more complex. In this scenario, the cooling does not result in notable improvement, and significant thermal effects persist. This is a clear indication that peak intensity-dependent nonlinear absorption mechanisms dominate, and likely disproportionately affect the center of the beam, making it more challenging to extract the heat. Two distinct timescales for thermal effects are likely responsible for heating in this regime: slow bulk thermal diffusion through the crystal, occurring over hundreds of seconds[110], and ultrafast nonlinear processes including two/multi-photon absorption[127], free carrier generation[128], and small polaron formation[129] that create per-pulse heating exceeding thermal management capabilities at high peak intensities.

It is important to also consider the regime of lower pump pulse energies and high average power at much higher repetition rates in the MHz region and above. In this regime, stronger focusing is needed to achieve sufficient peak intensities. When spot sizes become comparable or smaller to the interaction length in the crystal, walk-off and pump depletion effects that are average power independent become primary limiting factors[125]. Nevertheless, preliminary experimental studies at 13 MHz and using >100 W of average power show a clear reduction of the output yield at constant pulse energy with increasing duty cycle of the laser[130], indicating also additional limitations due to thermal effects, albeit significantly more moderate than in the case of high energies studied above. This finding can be conceptually reconciled with the general trends of the study above at higher energies: walk-off prevents focusing the beam to reach very high intensities, therefore, absolute peak intensities are lower, and therefore so are conversion efficiency but also heating effects due to nonlinear absorption.

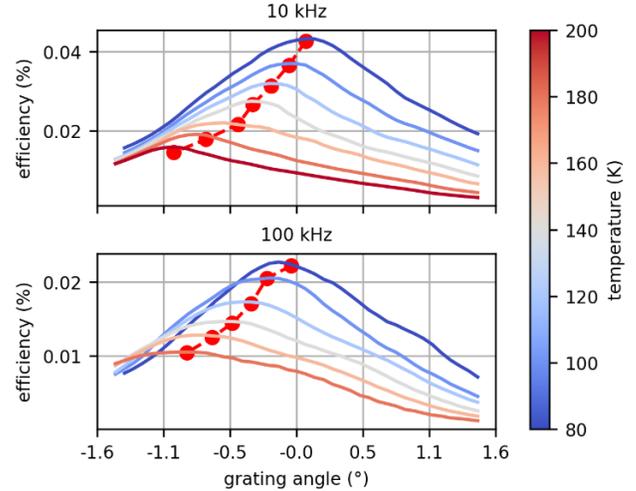

Figure 8: THz conversion efficiency versus grating angle at different crystal temperatures for 10 kHz (top) and 100 kHz (bottom) operation. Red circles show the optimal grating angles, which shift significantly with temperature. The broadening and shifting of efficiency curves demonstrate that thermal effects fundamentally alter phase-matching conditions rather than simply reducing conversion efficiency. A similar effect can be observed at a fixed cooling setpoint and with laser-induced heating when the pump power is increased.

This suggests that a very promising regime of operation for LN-TPF is that of intermediate pulse energies of hundreds of microjoules and hundreds of kHz repetition rate. This regime is now commonly accessible by commercial Yb lasers with tens of watts of average power. In this regime, enough pulse energy is available to circumvent the above mentioned walk-off limitations, and it was shown in Ref.[123] that no significant thermal effects are observed in LN-TPF at room temperature. In fact, several research groups have successfully implemented systems in this regime to achieve very high conversion efficiency THz sources with tens of mW of average power[35].

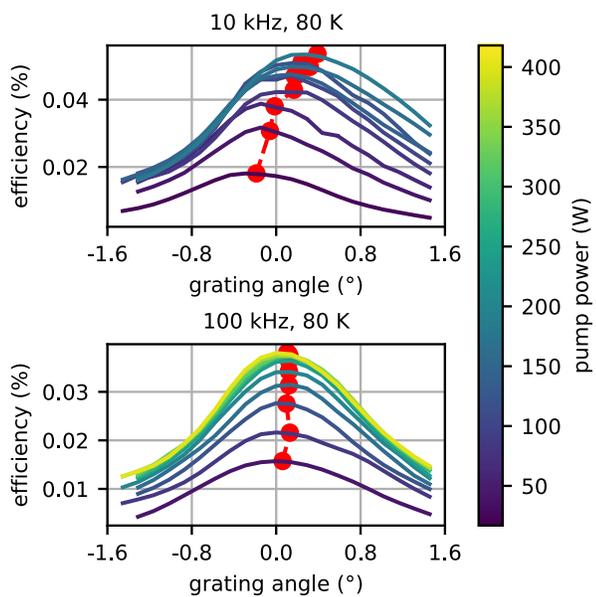

Figure 9: THz conversion efficiency versus grating angle at 10 and 100 kHz repetition rate with improved cooling geometry maintained at 80 K. Red circles show optimal grating angles, which remain stable across all power levels, demonstrating that improved thermal management eliminates grating angle drift at 100 kHz operation. The absence of such stabilization at 10 kHz suggests peak intensity-dependent nonlinear absorption mechanisms dominate at lower repetition rates.

Future strategies

Although the understanding of LN-TPF at high average power has significantly advanced in last years, there are still many unknowns. In particular, the regime of very high average power and high pulse energies (>>mJ) remains the most challenging. The measurements shown above are indicative of the complexity of optimizing LN-TPF at high average power. One important point to note at this point is that the study presented above was performed at a fixed pump spot size, i.e. the fluence level was not adapted to the reduced energy level available at 100 kHz. This introduces an additional level of complexity in performing such comparative studies. Normally, stronger focusing should be attempted at lower energy to achieve constant intensity on the crystal. Intuitively, this would result in stronger heating effects, however if the crystal size can be reduced, better thermal extraction can also be expected. One important takeaway message from this study is that every repetition rate and pulse energy range requires a separate optimization strategy, dedicated simulations, and a (different) heat removal strategy.

To fully understand limitations and develop optimal power scaling and specifically heat elimination strategies, more dedicated studies are required that combine time-resolved pump-probe spectroscopy to differentiate between short time-scale effects and delayed polaron formation processes occurring over ~100 fs timescales, wavelength-dependent nonlinear absorption measurements to map the spectral dependence around 1030 nm pumping, and systematic investigations of the interplay between crystal geometry, thermal transport, and peak intensity effects. From our current understanding of limiting effects at hundreds of watts of excitation average power, the regime of few mJ at repetition rates of >100 kHz to few MHz remains the most promising route to scale to the watt level using LN-TPF with considerable high THz fields. However, for applications where absolute THz pulse strength might not be the main criteria, the MHz repetition rate regime with moderate pulse energies (>100 µJ) is also a promising route.

There is also other practical difficulty in transitioning to higher repetition rates at fixed high pulse energies, which from the studies above is shown to require larger beams to accommodate the additional accumulation effects. Large beams at high average power are typically imaged through a grating, which is known to lead to aberrations in the pump beam and a decline in THz beam quality[131]. At high average power, this effect will therefore occur at lower pump pulse energies. Additional thermal aberrations and depolarization effects in all components in the setup also affect imaging. Consequently, alternative concepts that circumvent traditional telescope imaging are gaining popularity, particularly those employing echelon mirrors in place of a grating to induce a pulse front tilt[35,132]. Additionally, the introduction of a grating structure into lithium niobate itself, either on the front[133], or on the back[134] were explored as an option at low average power, but the hard and brittle structure of the material renders this approach challenging[135]. Nevertheless, explorations of such methods at high average power can be expected in the near future.

For much higher repetition rates of tens of MHz and beyond, a path to circumvent walk-off in LN-TPF scheme with small pump spot sizes is the so-called Cherenkov scheme in a slab. In this scheme, the pump laser beam is focused to a size smaller than the terahertz wavelength in one (focusing to a line) or two (focusing to a spot) directions, thus avoiding destructive interference of terahertz waves emitted by different parts of the crystal. This focusing allows to achieve high optical intensity, which is required for efficient nonlinear optical rectification, with lower energy laser pulses than in the tilted-pulse-front configuration. Such a scheme could additionally offer benefits in terms of thermal removal. One possible drawback is the requirement for a contacted Si prism for output coupling the THz beam, which imposes constraints for thermal management of the LN-Si interface.

We also note that we have experimentally observed significant differences between crystals from different vendors, which further highlights the need for coordinated and continuous collaboration with nonlinear crystal growth researchers and manufacturers.

In summary, over the past several decades, LN has undergone significant advancements in average power, primarily driven by Yb-based systems. We can expect further improvements, on the one hand, through thermal engineering efforts and, on the other hand, through new concepts to avoid aberrations that become increasingly important at high average power. It can be expected that multi-W-level sources will become available in the near future, when moderate pump pulse energies are used at high repetition rates.

### IV.a.ii Organic crystals

#### General considerations

Another family of materials that has attracted significant attention for optical rectification are nonlinear organic crystals. The main advantage of these materials is a comparatively high second-order nonlinear susceptibility, usually one order of magnitude higher than the corresponding values for inorganic materials (Table 2). Another key advantage of these crystals is their very broadband group-to-phase velocity matching, often achieved collinearly or with reasonable crystal thickness at near-infrared wavelengths[105,136]. These favorable properties allowed these sources to gain popularity for generating strong fields and broad bandwidths, with many developments relying on high-energy Ti:sapphire laser systems at a central wavelength of 800 nm[106,137,138] and other systems at longer wavelengths of 1950 nm and 3900 nm to exploit advantageous velocity matching properties while avoiding multiphoton absorption (MPA)[139]. However, most studies were performed at very low repetition rates. In fact, the main drawback of these materials is their poor thermal properties: very low thermal conductivity (Table 2), low melting temperature, and low damage threshold. This means special attention needs to be paid to thermal management within these critical conditions when considering the use of high average powers. We detail below the attempts made with temporally compressed Yb-lasers to benefit from the broadband emission[86]. We note that compared to the above-mentioned crystals, organic crystals have so far been explored with driving average powers levels in the order of few watts to tens of watts, leaving significant space for future work.

#### State of the art using Yb-based lasers

The first high-repetition-rate results using organic crystals were reported in ref.[140] and demonstrated mW-level, few-cycle THz pulse generation, pumped by a 2.5-W average power, 10-MHz repetition rate laser system. A Yb-doped fiber laser, incorporating external pulse compression via a large-mode-area solid-core photonic crystal fiber, was used to pump HMQ-TMS (2-(4-hydroxy-3-methoxystyryl)-1-methylquinolinium 2,4,6-trimethylbenzenesulfonate). A maximum conversion efficiency of 0.055% was achieved—an order of magnitude higher than that of inorganic nonlinear crystals such as GaP under comparable pump conditions, i.e., wavelength and repetition rate. For thermal management, this result utilized optical chopping with a frequency of 673 Hz, which proved essential at these high repetition rates to prevent thermal damage.

Following this first demonstration the more widely available BNA (N-benzyl-2-methyl-4-nitroaniline, Fig. 10(a)) was investigated using a 13-MHz high-power Yb-based thin-disk oscillator, nonlinearly compressed to sub-100 fs[141]. BNA is not ideally velocity matched at 1030 nm; however, the use of thin crystals still yields significant bandwidth and high efficiency at 1-µm pumping. Using a diamond-heatsinked crystal, mW-level output was achieved. In this result, it was demonstrated that saturation in conversion efficiency in this regime is driven by the pump average power rather than pump pulse energy, confirming that thermal effects are the primary limiting factor. These results are shown in Fig. 10(c), adapted from Ref.[141], showing the THz power and THz conversion efficiency versus duty cycle (illustrated in Fig. 10(b)) of the optical chopper at constant pump pulse energy. The THz-to-optical conversion efficiency is expected to be continuous, as the pump pulse energy remains constant. However, the measurement shows a drop in efficiency due to strong thermal accumulation effects at MHz repetition rate. In this case, burst-mode pumping is essential to apply high average power levels. In fact, when the time between pulses (e.g., 100 ns at 10 MHz) becomes significantly shorter than the thermal relaxation time of the crystal (typically in the ms range[142]), the most appropriate metric for calculating the damage threshold is the low repetition rate burst, determined by the cumulative energy of the femtosecond pulses during the open time of the chopper wheel cycle, rather than the energy of a single pulse. When using the burst energy as a metric, the damage thresholds measured were comparable to those measured using low-repetition-rate nanosecond-pulsed systems, showing clearly the dominant thermal effects.

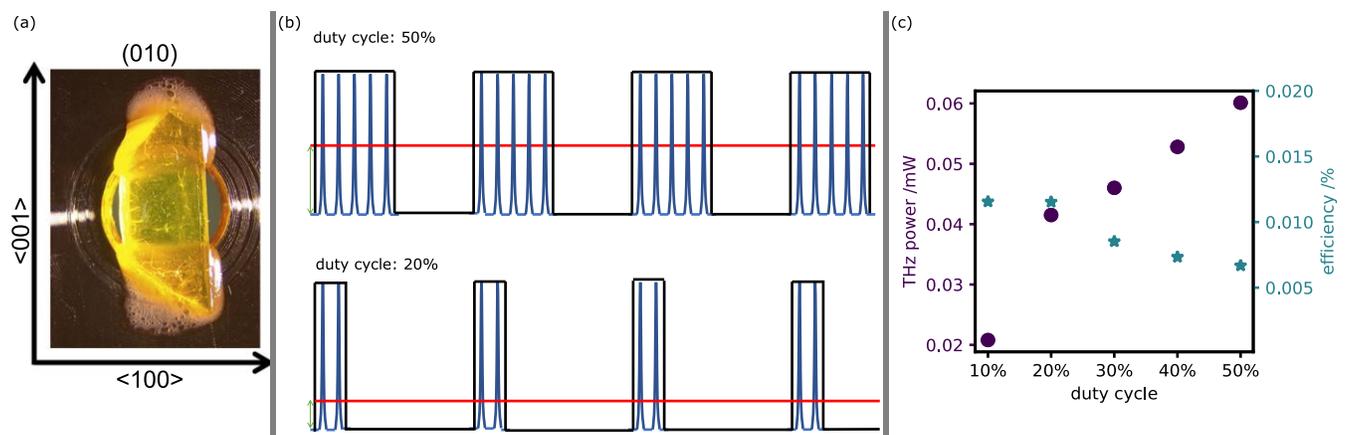

Figure 10: (a) Picture of the BNA crystal. (b) Illustration of the duty cycle for effective repetition rate reduction in burst-mode operation. (c) left axis: measured THz average power using a pyroelectric detector, right axis: calculated optical-to-THz conversion efficiency. Figures adapted from Ref.[177].

In a further study, active cooling of these crystals was investigated, and it was confirmed that optical chopping is significantly more efficient to thermally manage the crystals, due to their low thermal conductivity[143]. While cryogenic cooling did not affect the damage threshold of the crystals, high-frequency components (>2 THz) exhibited significantly reduced absorption, consistent with the temperature dependence of the THz absorption properties of the crystal.

Further experiments with BNA pumped with Yb-based lasers were demonstrated in Ref. [123] at a slightly lower repetition rate of 400 kHz. In this case, the inter-pulse time is still significantly shorter than the thermal relaxation time of the crystal; therefore, heat accumulation effects remain, and optical chopping remains essential. However, this regime is considerably easier to optimize for high efficiency, since looser focusing conditions still yield good conversion efficiency due to higher available pulse energies, leading to lower operation temperatures in the crystal. Although repetition rates >100 kHz are significantly lower than the MHz region considered above, fast mechanical scanning methods can already be considered for detection and still accelerate measurement times compared to previous uses of these crystals. In Ref.[85], a THz average power of 5.6 mW with an efficiency of 0.12% was obtained at 4.7-W pump power (at a duty cycle of 50%) at 400 kHz.

More recently, advancements in new organic crystals and their growth, as reported in Ref. [144], have also enabled us to use other crystals in combination with Yb lasers such as high-quality MNA (2-Amino-5-Nitrotoluene). A recent study demonstrated a THz-TDS based on this crystal driven by a commercial Yb-based laser operating at a 100-kHz repetition rate[38]. Under optimized conditions, an average THz power of 11 mW was obtained with 8 W of driving power on the crystal at a chopping duty cycle of 50%, which to the best of our knowledge is the highest power obtained with collinear optical rectification in the THz region so far. Remarkably, the excellent quality of the crystal made broadband detection, using the MNA crystal, as well possible: the THz-TDS demonstrated broad spectral bandwidth extending up to 9 THz with a DR exceeding 40 dB across most of the detected bandwidth within just 5 minutes of measurement time. Broadband extraction of the complex refractive index was demonstrated for several materials over the entire bandwidth of up to 8 THz in a few-minute measurement time. This exceptional performance was accomplished by utilizing MNA not only as a high-power THz emitter but also as a broadband and highly sensitive detector. This approach was previously unfeasible due to the typically poor surface quality of organic crystals and strong birefringence of the crystals, pointing to more advances in the future as new crystals with improved crystal qualities become available.

Forward-looking developments

One of the primary challenges in scaling average power in organic crystals remains heat management, particularly for very high repetition rates, i.e. >>MHz. Low-duty cycle pumping to match the off-time of the laser with the thermal relaxation time of the crystals is possibly the most efficient alternative for cooling, however this can become a constraint for laser systems without natural burst-mode operation, as the ratio of required laser power to incident power becomes very large – thus effectively again reducing overall conversion efficiency. Furthermore, detection schemes that rely on high-frequency modulation for noise reduction become challenging to realize at very low duty cycles. More sophisticated heatsink cooling geometries, such as sandwiched diamond or silicon carbide (SiC), can also be considered in more detail to partially alleviate these issues[145].

Another challenge that remains a topic of exploration is the variability in crystal quality and the long-term degradation of the crystals. Although the first tests are presented in Ref. [38] (supplementary materials) for MNA, long term studies where the crystals are irradiated for many days are still required, particularly at high average powers and in varying conditions (focusing, duty cycle, etc.). These aspects contribute to difficulties in source design and reproducibility.

With the increased availability of lasers with higher average power and pulse energy simultaneously, the need for larger organic crystal sizes with higher surface quality and a higher damage threshold will continue to increase. Advanced research on finding new organic crystals[146] is also critical in advancing this area. Through data mining, new crystals are discovered, synthesized, and characterized.

Other studies[147] present design strategies for ionic organic THz crystals, focusing on suppressing intrinsic phonon vibrations to achieve low THz absorption, good phase matching between optical and THz pulses, and improved bulk single-crystal growth ability with high optical quality and large-area plate-shaped morphology.

Finally, high-power THz emitters based on organic crystals will immensely benefit from ongoing efforts in broadband THz detection at high repetition rates, since in spite of exceptional performance, the overall TDS systems are often bandwidth-limited by the detection method used. We highlighted above a first very promising result using organic crystals for EOS for the first time. Other approaches for broadband detection will be discussed in the Section V.a on fast and sensitive time-domain detection, which can be combined with THz generation based on organic crystals.

## IV.b Gas-plasma sources

General considerations

When considering applications where ultrabroad THz bandwidths >>10 THz are required, using gas as a nonlinear medium[148,149] is an increasingly attractive alternative, as it eliminates phonon absorption peaks observed in crystals and semiconductors. For average power and peak power scaling, gases do not suffer from damage as crystals do.

The most commonly exploited generation mechanism of THz frequencies using gases is photoionization and linear acceleration of free electrons in an asymmetric field in a gas, most commonly achieved by two-color excitation in air filament plasma. The first demonstration of this technique dates back to 2000[150], and has since then seen significant progress: ultra-broadband THz pulses exceeding 100 THz are demonstrated, and these sources can routinely reach MV/cm peak fields[150,151]. A recent summary of the state-of-the-art of these sources is given in Ref. [152].

The primary challenge of this technique is the requirement for extremely high laser driving peak intensities to photoionize a substantial fraction of the gas. So far, therefore, the vast majority of experiments in this area have been restricted to kHz repetition rate systems and below, significantly limiting the applicability of these otherwise unique sources, due to their low acquisition speeds and low SNR. This is exacerbated, as already explained for the organic crystals above, by difficulties in detecting these ultra-broad bandwidths with standard methods. High-power Yb lasers make it possible to extend the operation of these plasma sources to high repetition rates, opening the door to more wide and practical implementation of these ultrabroadband sources.

State-of-the-art using Yb-based lasers

The first Yb-fiber driven gas-plasma THz source achieved 50 mW of average power[153]. This source was driven by a 16-channel coherently combined fiber amplifier with post-compression in a hollow-core capillary. The two-color gas plasma was generated by 30-fs, 1.2-mJ pulses at a 100-kHz repetition rate in neon gas. This result was followed by order-of-magnitude scaling to 640 mW within one year[41], this time using 37-fs, 1.3-mJ driving pulses at a repetition rate of 500 kHz. In these results, it was confirmed experimentally that using static gas at high repetition rates reduces conversion efficiency due to hydrodynamic effects that affect the gas density and thus ionization efficiency[40,154,155], but no quantitative studies were performed. Power scaling of the THz output could only then be achieved using a high-pressure gas jet for gas replenishment[41].

In a different study at slightly lower average power[154], the authors reported reduced efficiency at 50-kHz repetition rate in air, however no detailed study of the hydrodynamics was performed. More recently, the effect of air hydrodynamics using filaments has been studied and quantified in greater detail. In Ref.[156], it was shown via interferometry that single-color filament plasmas with repetition rates of 40 and 100 kHz driven by multi-mJ sub-picosecond pulses result in quasi-stationary density depressions of 8% and 18%, respectively. In a following study using lower pulse energy (Fig. 11), but this time a broader repetition rate region, two important findings were made: on the one hand, the conversion efficiency loss to the THz was very moderate and only a small reduction of 19% (in nitrogen, 31% in ambient air) was observed when going from 1 to 100 kHz[40]; on the other hand this could only be achieved by optimizing the relative phase between the fundamental and second harmonic as well as the collection geometry of the THz light at each repetition rate. In fact, a shift of the effective focal position of the filament occurs due to the hydrodynamics[155,157]. An average power of 2.8 mW at 100-kHz repetition rate was obtained at a conversion efficiency of $1.45 \cdot 10^{-4}$ in ambient air (Fig. 12). Notably, this value is comparable to those typically observed in the low repetition-rate regime, although the peak power of the pump pulses remained below the critical peak power threshold for air of about 6 GW[158] (Fig. 11(c)). The broad spectral bandwidth of the pump pulses achieved by post-compression (Section III.b) enabled the generation and detection of an ultra-broadband spectrum extending up to approximately 23.5 THz.

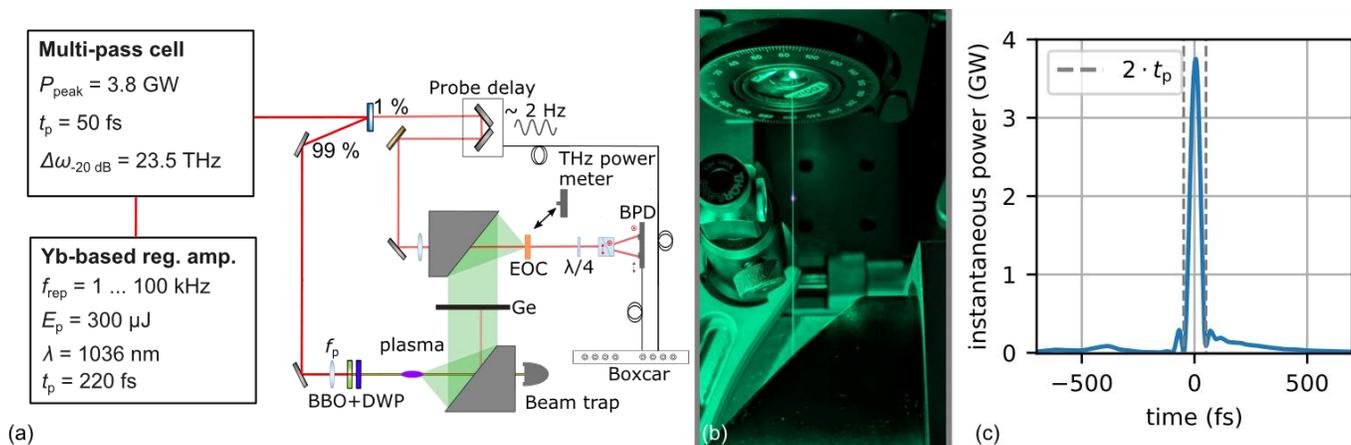

*Figure 11: (a) Two-color air-plasma THz generation/detection setup for a high repetition-rate Yb-based regenerative amplifier. BBO: β-barium-borate, DWP: dual-wavelength waveplate, EOC: electro-optic crystal, BPD: balanced photo detector. (b) Picture of the two-color air plasma. (c) Time-dependent instantaneous power of the nonlinearly compressed pump pulses. (a), (b), and (c) adapted from Ref [40].*

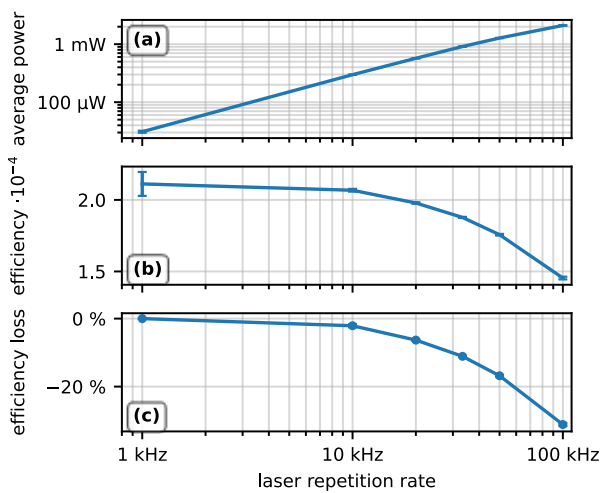

Figure 12: Results adapted from Ref.[40] in air. THz average power for repetition rates between 1 and 100 kHz. The pump pulses with 3.8 GW of peak power were focused through a 100-mm focal-length lens. The second harmonic was generated collinearly in the focusing beam in a 100-µm thick β-barium borate crystal.

Forward-looking developments

An attractive area of future development will be to expand plasma-based THz sources to even higher repetition rates. For much higher repetition rates of several MHz and beyond, micro-plasma-type sources will become relevant[159], but the above-mentioned accumulation depletion effects will become significantly more severe. While improved gas flow engineering can eliminate density depletion effects at kHz repetition rates, fundamental limitations emerge as shorter inter-pulse separations demand ultrasonic laminar gas flows, which are only achievable for minimal gas volumes. Burst-mode operation is another potential path for thermal management of the gas, like measures applied in organic crystals[136].

Other approaches that can also help mitigate thermal effects are advanced spatial arrangements of individual filaments when operating significantly above critical peak power. Plasma sheets[160] and vortex beams[161,162] also reduce local heat deposition while ensuring single filamentation, enabling ultra-broadband THz transients from pump pulses far exceeding critical power through advanced THz imaging techniques. Real-time adaptive optics systems that dynamically compensate for hydrodynamic effects and phase slippage represent the most sophisticated approach to optimizing THz generation in gases, potentially enhancing performance even in turbulent environments for free-space remote sensing applications[163]. Furthermore, advanced control of the THz emission can be achieved by engineered spatial or temporal chirps. Very recently, it was demonstrated that chromatic aberrations of a pre-chirped pump pulse allow for controlled steering of the THz emission angle from an air plasma[164].

Alternatively to gas targets, the first experiments in liquid media for plasma-driven THz sources showed potential due to the lower ionization threshold which were close to those of solids[22,165,166]. A double pump scheme has been shown to enhance the THz yield from a liquid sample by nearly an order of magnitude, which the authors attributed to the ionization mechanism[167], however significant work remains to understand scaling principles before high repetition rates become relevant.

Last but not least, excitation by pulse trains with optimized temporal spacing might be an alternative to engineer optimized plasma characteristics for maximizing the optical-to-THz conversion efficiency in both liquid and gaseous targets[168–170]. Besides the abovementioned possible path towards tailored plasmas driven by sources operating at several to hundreds of kHz, as lasers with > 10 kW become available[73], ultra-high repetition rates exceeding 100 MHz will also need to consider plasma chemical accumulation effects.

## IV.c Photoconductive emitters

General considerations

So far, all techniques considered rely heavily on pulse peak intensity to drive THz generation, either through the $\chi^{(2)}$ response of nonlinear crystals or through the need for photoionization. This restricts these methods to pulsed systems that are rather complex, typically requiring CPA. Photoconductive antennas (PCAs) operate on the principle of acceleration of carriers in a biased semiconductor. As such, the requirements for driving peak intensity are significantly reduced compared to nonlinear crystals since the role of the optical pump is to modulate the conductivity of the semiconductor, which has very high absorption cross sections. The acceleration energy and thus THz emission originates from the bias voltage, and the efficiency is not determined by the Manley–Rowe limit where one near-infrared photon is required for a THz photon, making this a very attractive method to go to very high repetition rates and very compact and cost-effective laser systems.

Photoconductive emitters are the most widely used sources and detectors for linear THz-TDS applications, with their operation principles being well-understood and main scaling principles treated in literature[171,172]. In particular, they represent most commercial TDS systems deployed in industrial applications.

In the context of potential high average power excitation such emitters are very promising for average power scaling at very high repetition rates at 100 MHz and beyond, which remains a relatively unexplored region (see Fig. 3), mainly due to the moderate driving energies and peak powers that remain insufficient for the techniques explained above.

At these ultra-high repetition rates, SNR and DR in the final TDS system are driven mainly by averaging rather than by single-pulse strength; therefore, the amplitude and phase noise properties of the driving laser and the THz pulse train become a more important criteria than in previously presented cases. Although low-noise lasers are commonly available at low average power high-average-power systems typically exhibit higher noise levels due to requirements for active cooling and high-power pump diodes. A few exceptions have been demonstrated but have focused on applications other than THz generation[173–175], pointing to an area of future research for laser development for the specific needs of THz-TDS. We note also that PC-based emitters can be modulated at extremely high frequencies, therefore also alleviating some of these low noise requirements in critical frequency regions [176,177].

The progress of PC emitter performance has therefore closely followed advances in high-repetition-rate laser systems with low noise levels. Mode-locked Ti:sapphire lasers operating around 800 nm (photon energy ≈ 1.55 eV) with typical repetition rates of 80 MHz were, in this context, an outstanding match to GaAs PCAs, with a bandgap of 1.42 eV. For market deployment, new PCA materials based on InGaAs were developed for excitation with Erbium fiber lasers at telecom wavelength of 1.55 μm (photon energy 0.8 eV)[178,179]. Since then, various strategies, such as ion implantation to reduce carrier lifetime and low-temperature-grown InGaAs, have been employed to enhance performance under fiber-laser excitation[50,52,180], primarily aimed at maximizing DR by improving detection sensitivity, rather than increasing average power. Nowadays, Erbium-laser driven InGaAs-based PCAs are an integral part of modern commercial THz TDS systems, with state-of-the-art PCAs reaching THz powers of 958 μW with 55 mW of optical power[181]. Other concepts, such as plasmonic-enhanced GaAs-based PCAs, have demonstrated 4 mW of THz power with 720 mW of excitation power using a Ti:sapphire laser system[180]; and photoconductive connected arrays (PCCA), which integrate an N × N PCA matrix with a microlens array to focus the pump light directly onto the active area, have been explored. This geometry has yielded 712 µW of THz power with 500 mW of laser power at 800 nm[182].

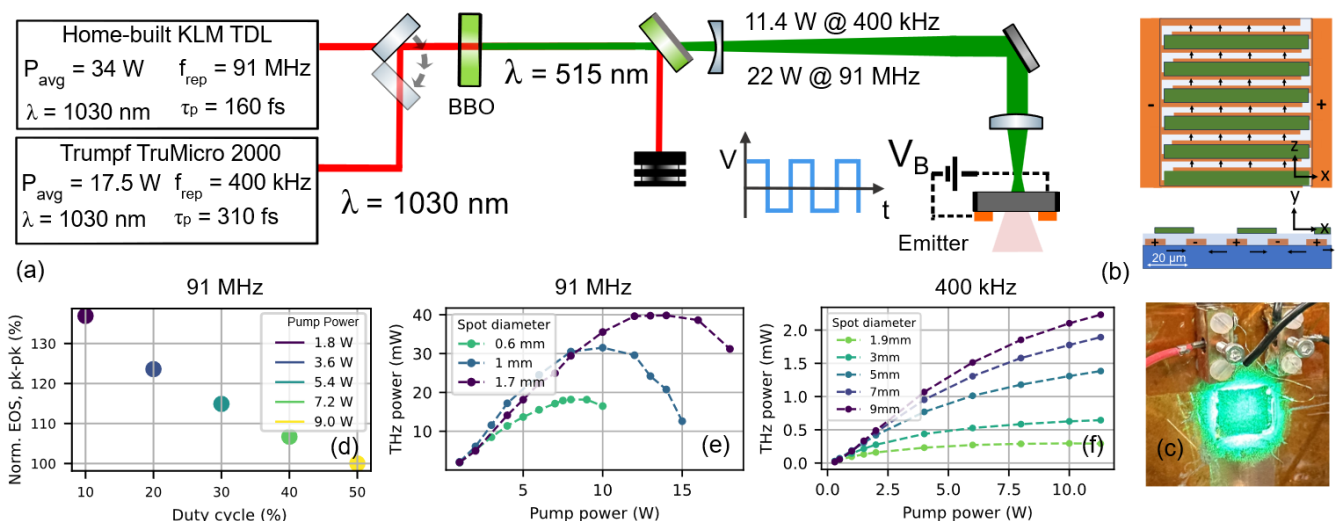

Figure 13: (a) General schematic of setups so far explored for high power LAE with frequency doubled Yb lasers at 400 kHz and 91 MHz (b) Geometry of microstructured LAE from top and side view. (c) LAE in water-cooled copper mount, irradiated with green laser. (d) normalized peak-to-peak value of the EOS signals in the time domain versus each duty cycle of the optical chopper for case of 91 MHz repetition rate. Legend shows the incident optical average power on the LAE for each corresponding duty cycle of the optical chopper. (e) THz power versus pump power for various laser spot size on the LAE at 91 MHz excitation. (f) THz power versus pump power for various laser spot sizes on the LAE at 400 kHz excitation. Figures adapted from Ref.[29].

The average power scaling of standard PC emitters is challenging due to the small PC gap dimensions of a few µm$^2$ and extremely efficient pump absorption, which prevents the application of high average powers without strong thermal effects and damage. Increasing the electrode gap size is one way to scale excitation spot areas and power[183,184]. However, this then requires kV bias voltages. A more promising approach is to use microstructured interdigitated large-area photoconductive THz emitters (LAEs)[30,185,186]. LAEs operate on the same principle and have significantly larger active areas (1–100 mm$^2$) with microstructured electrodes of much smaller dimensions than the THz wavelength that provide high acceleration fields in the kV/cm range at moderate bias voltages of a few tens of volts. Early results with GaAs LAEs using Ti:sapphire amplifier drivers achieved 1.5 mW of THz power at 250-kHz repetition rate with an excitation power of 800 mW[186], limited by the average power of the laser.

Scaling the average power applied to these structures is a rapidly emerging area. However, the photon energy at 1030 nm is too small for the bandgap of GaAs LAEs. Some experiments were reported making use of defects in GaAs, but the devices operated at very low efficiency[30]. Bandgap engineering of the InGaAs platform developed for 1550 nm can be extended to 1030 nm as has been shown for receivers[51]; however, the entire technology platform needs to be re-demonstrated and optimized. This is why, so far, less progress has been made in this area, indicating further areas for development. Nevertheless, a path to partially circumvent this problem and explore average power scaling is to frequency double Yb lasers (photon energy 2.4 eV at 515 nm) and utilize existing LAE technology based on GaAs, albeit with some differences in the scaling laws due to the relatively large photon energy compared to the bandgap. This approach was recently explored using high-average-power Yb lasers, and we present the first studies and findings below.

<u>State-of-the-art using Yb-drivers</u>
The first attempts to average power scale of LAEs were performed using a high-power Yb thin-disk oscillator at 91-MHz repetition rate with up to 34 W of average power, frequency doubled to 515 nm[31] (18-W average power available with 100-fs pulses), to excite a 10 × 10 mm$^2$ interdigitated metal-semiconductor-metal (MSM) LAE based on semi-insulating GaAs (SI-GaAs), fabricated at the Helmholtz-Zentrum Dresden-Rossendorf[183]. Figure 13(b) shows the LAE structure from the top and side views. Electrode patterns are indicated with an orange color, and the arrows show the bias field direction. The MSM structure is masked by a second metallization layer (green pattern) isolated from the MSM electrodes. The isolation layer on MSM blocks the excitation laser beam in every second period of the MSM structure. In this way, charge carriers are excited and accelerated unidirectionally, and the THz fields radiated interfere constructively in the far field. Figure 13(c) shows that a 10 × 10 mm$^2$ LAE is housed in a water-cooled copper mount that extracts a small fraction of the accumulated heat from the emitter sides. In these experiments, it was demonstrated that the thermal load from the optical power was the dominant factor limiting THz conversion efficiency, and that further power scaling requires improved thermal management. Figure 13(d) illustrates this by showing THz yield as a function of duty cycle of the optical chopper (from 50% to 10%). By decreasing the duty cycle of the chopper, corresponding to a decrease in optical average power but no change in the carriers generated by each pulse, the peak-to-peak value of the EOS signal increased by about 37%. Figure 13(e) shows THz power versus pump power as a function of spot size. By increasing the spot size on the LAE, saturation occurs at higher pump power due to spreading the heat on a larger area, leading to improved LAE efficiency. However, further enlarging the pump spot reduces efficiency again, due to a too low fluence per pulse. Under optimized conditions, with 9 W of pump power (at a 50% duty cycle) and a 1.7-mm spot diameter, a THz power of 65 µW was achieved, indicating very limited conversion efficiency. In these experiments, it was also shown that the LAE can withstand 18 W of optical power without degradation.

In a continuation experiment, the 400-kHz repetition rate regime was explored, providing higher single-pulse energy while maintaining relatively high average power. The higher energy per pulse is beneficial in this case, as it generates higher carrier densities at larger spots, and the repetition rate remains high enough for fast acquisition and averaging[29]. The same LAE structure as in the previous experiment was used, excited this time by a frequency-doubled Yb-based commercial laser amplifier that delivers up to 11.4 W of average power at 515 nm at a repetition rate of 400 kHz. In Figure 13(f) shows the generated THz power versus pump power as a function of spot size. For all beam diameters, increasing pump power leads to higher THz power, and also the efficiency significantly improves compared to 91 MHz excitation, indicative of a decreased thermal load at lower repetition rates. In this way, substantially higher power and conversion efficiency was achieved, reaching 6.7 mW of THz output, marking the highest average power demonstrated so far with PC emitters, albeit still a moderate conversion efficiency of $3.8 \cdot 10^{-4}$ (IR-to-THz) compared to standard PC emitters[29]. A significant limitation of the semi-insulating GaAs MSM LAEs used in our experiments is that the metallization layer covers ~75% of the active area[31,187]. Considering this aspect and Fresnel reflection losses at the air-GaAs interface, only about 20% of the pump power is used to excite the photoconductive material, highlighting the advantage of the PCCA approach.

In our very recent development, 10 mW of average power was obtained using 19 × 19-element array (1.9 × 1.9 mm$^2$) with the same laser system as Ref.[31] at 91 MHz. This was achieved with 10 W of pump power (at 50% duty cycle) using 515-nm pulses and a 1.9-mm spot diameter. These results represent the highest THz average power ever achieved with a PC emitter at MHz repetition rate, to the best of our knowledge[188].

Forward-looking developments

One of the main drawbacks so far in the experimental realization is the need to frequency double the Yb driving laser. This not only adds experimental complexity and reduces overall system efficiency, but also the photon energy is not well matched to the bandgap of GaAs (2.4 eV at 515 nm, bandgap 1.4 eV[189]). The excess photon energy contributes to additional heating and leads to transfer of photo-generated carriers with high probability to L-valley, where carrier mobility is significantly reduced (effective mass ~30 times higher than Γ-valley)[31], reducing THz generation efficiency. In future devices, alternative materials optimized for direct excitation at 1030 nm, the fundamental wavelength of Yb lasers will be key in achieving optimal operation. In this direction, ErAs:InAlGaAs-based PCA receivers optimized for 1030 nm[51] have recently been shown, and the material can in principle be adapted for LAEs.

Thermal management, particularly arising from optical heating, remains a bottleneck in particular for very high repetition rates >>MHz, where very large spots cannot be used efficiently. Beyond careful material selection and electrode design, heat dissipation can be improved by designing PC emitters with transparent heatsinks and/or in reflection geometry, implementing backside cooling schemes. Further improvements in thermal management and the development of new semiconductor materials compatible with Yb-based laser excitation can be expected to support further progress in coming years.

Another critical area of research to support the application of these sources is to optimize overall SNR in connection with PC receivers, such as those developed in the context of 1030 nm lasers[51]. To achieve this goal, the noise properties of high-power lasers need to be more carefully evaluated and addressed with active stabilization measures, in particular in critical frequency regions that affect SNR in a given detection scenario.

## IV.d Spintronic THz emitters

General considerations

Spintronic THz emitters have recently attracted attention for applications requiring broad, gapless bandwidths, but without the requirements for high peak-intensity pulses in air/gas plasma sources (Section IV.b). Their generation mechanism exploit ultrafast spin-to-charge current conversion[23,25]. The physical mechanism relies on femtosecond laser excitation of ferromagnetic (FM) layers in FM/nonmagnetic (NM) heterostructures, where spin-polarized electrons are promoted to higher energy bands and subsequently diffuse as spin currents into the adjacent NM layer[26]. The inverse spin Hall effect (ISHE) in the NM layer converts this longitudinal spin current into a transverse charge current, generating THz electromagnetic radiation.

The key advantage of spintronic THz emitters lies in their ultrabroadband, gapless emission spectrum and the fact that they do not rely on high peak intensity, but rather on fluence to operate efficiently. While emission extending to 30 THz has been reported through deconvolution techniques[190], practically usable bandwidth is typically limited to approximately 6 to 11 THz, often limited by detection constraints rather than the emitter itself[27,28]. Another advantage is their wavelength-independent response. Efficient THz generation has been demonstrated across excitation wavelengths from 400 nm to 1550 nm, with conversion efficiency remaining essentially constant[191–193], which contrasts strongly with all other generation mechanisms.

State-of-the-art using Yb-drivers

With regards to average power scaling, spintronic THz emitters present complex challenges that have only recently started to be explored. Experimental observations reveal power saturation behavior that involves two distinct mechanisms: one related to the steady-state temperature of the emitter scaling with average power, and another scaling with fluence or peak power, possibly involving transient heating of the metallic films[28]. The nanometer-thin magnetic layers exhibit low damage thresholds under focused excitation conditions, with permanent damage occurring at fluences beyond 5 mJ/cm$^2$, limiting many power scaling approaches[194]. On the other hand, the thin metallic layer is ideal for cooling, therefore strategies for heat removal can partially alleviate these limitations, for example via backside cooling[28].

One additional challenge is the generally weak optical-to-THz conversion efficiency ranging from $10^{-6}$ to $10^{-4}$, lower than optimized nonlinear crystals or photoconductive emitters[195]. The intrinsic weak ISHE mechanism requires optimization of both the spin current generation and electromagnetic extraction processes. While various photonic enhancement approaches for improving conversion efficiency have been demonstrated at 800 nm, these require adaptation for 1030 nm operation and will be discussed in the forward-looking section. However, significant advances in THz field control have

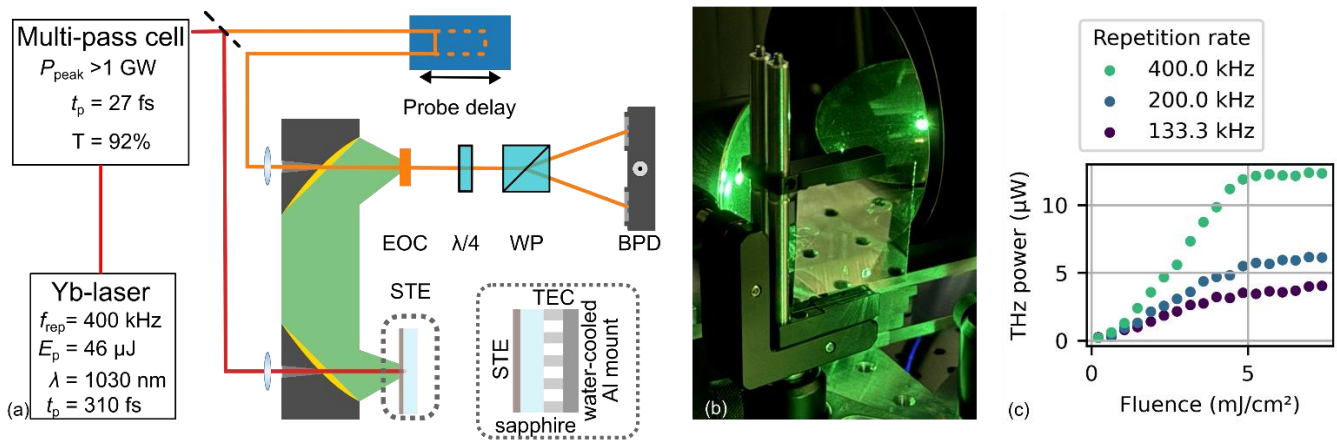

*Figure 14 (a) THz time domain spectroscopy setup for backcooled Spintronic THz emitter (STE), pumped by a high repetition-rate Yb-based laser system in connection with a multi-pass cell (MPC). EOC: electro-optic crystal, WP: Wollaston Prism, BPD: balanced photo detector. (b) Picture of the STE. (c) Saturation fluence of approximately 5 mJ/cm^2 is independent of repetition rate. (c) adapted from Ref. [39].*

been achieved at 1030 nm. Metasurface integration enables controlling polarization and chirality of THz output, with excellent circular polarization achieved over broad terahertz bandwidths[100]. Programmable spintronic emitters using exchange-biased heterostructures allow flexible generation of structured terahertz waves through precise magnetization pattern programming, enabling diverse polarization states[196].

Average power scaling using Yb-lasers up to 7 W of average power at 400 kHz was systematically explored[28]. In this result, a reflection geometry with active backside cooling was used to reduce temperature rise from 16.3 K/W (free-standing) to 8.1 K/W and enabling 39 µW THz output at 7 W pump power with conversion efficiency of $5.6 \cdot 10^{-6}$. Whereas thermal management helps to prevent damage, interestingly, it was shown that the devices have an optimal (linear) operation fluence of around 1 mJ/cm$^2$ that does not depend on the average power. Typical onset of saturation is reached at approximately 5 mJ/cm$^2$, as it can also be seen in Fig. 14(c). Thermal management allows to pump the devices further into saturation but without improving THz average power.

More recent work by Vaitsi et al. addressed thermal limitations through mechanical rotation of the spintronic emitter at several hundred Hz, distributing absorbed power over larger areas, and enabling unprecedented operation at MHz repetition rates with microjoule pulse energies[27]. Using an 80-W Yb-laser system, they achieved 10 kV/cm peak THz fields with power densities up to 925 W/cm$^2$ while maintaining optimal fluence conditions. Remarkably, no instabilities or performance degradation in the TDS were observed due to the mechanical rotation, and such a high repetition rate system enabled applications in THz scanning tunneling microscopy that require high repetition rates and stable broadband operation[197].

Despite modest conversion efficiencies, the combination of gapless ultrabroadband emission, wavelength flexibility, and exceptional system integration capabilities position spintronic THz emitters as a distinct class of THz sources. The system integration advantages are comparable to those of photoconductive emitters and arise from fundamental design simplicity compared to conventional THz sources. Their nanometer-scale thickness enables deposition on virtually any substrate without mechanical constraints that often limit bulky nonlinear crystals. The absence of velocity-matching requirements eliminates the precise angular alignment, temperature control, and beam conditioning needed for crystal-based sources. Wavelength-independent operation from 400 nm to 1550 nm allows integration with diverse laser technologies without source-specific optimization, contrasting with photoconductive antennas that require wavelength-matched bandgaps. Standard thin-film fabrication enables direct integration into semiconductor processing workflows for on-chip applications. Demonstrated integration examples, though at different wavelengths, include fiber-tip integration for compact systems and combining spintronic THz emitter with an antenna structure[192,198]. These integration advantages make spintronic THz emitters particularly suited for applications where bandwidth, wavelength flexibility, and compact system design outweigh absolute power requirements[195].

Forward-looking developments

Field enhancement strategies represent one promising path for further performance improvements, with most demonstrations focused on 800-nm Ti:sapphire systems requiring adaptation for 1030-nm operation. THz resonant cavities using sapphire substrates with metallic mirrors have demonstrated up to 2× field enhancement through

λ/4 cavity resonance[199]. Photonic crystal structures employing multilayer metal-dielectric stacks have achieved 1.7× efficiency improvements by maximizing laser absorption in the metallic emitter layers[200]. Optimized high- and anti-reflection coatings using alternating $SiO_2$/$Ta_2O_5$ layers have shown 35% THz signal enhancement while eliminating transmitted pump light[201]. Even though these wavelength-specific approaches require redesign for 1030 nm Yb-laser systems, the fundamental enhancement concepts remain applicable for high-power applications.

Elimination of external magnetic fields would also address scaling limitations for large-aperture applications, i.e. for a regime of both high average power and high energy excitation. Antiferromagnetic-ferromagnetic heterostructures have demonstrated large-area emitters generating strong THz fields without external magnets[202]. This approach enables arbitrarily large STEs for high-energy applications, removing the constraint of maintaining uniform magnetic fields over areas required for kilowatt-level excitation. While initial demonstrations used Ti:sapphire lasers, the underlying physics suggests compatibility with Yb systems.

Material engineering approaches focus on optimizing the spin-to-charge conversion processes through targeted impurities and interface modifications. These wavelength-independent strategies have demonstrated substantial efficiency enhancements while maintaining the broad spectral response characteristic of STEs[203]. For example, MgO impurity doping in Pt layers achieved 2× improvements[203]. Such approaches offer immediate applicability to high-power Yb-laser systems without architectural modifications.

Thermal management beyond mechanical rotation[27] remains relevant for further scaling to kilowatt-class excitation. Diamond heatsinking the thin metallic layers, and even cryogenic temperature operation could be an interesting path to pursue at very high repetition rates, provided a minimum amount of fluence can be reached. Burst-mode operation synchronized with thermal relaxation timescales offers one pathway to exploit peak power advantages while maintaining thermal stability.

Integration with kilowatt-class Yb systems requires rethinking traditional STE operation parameters. Rather than maximizing fluence within damage thresholds, future approaches may optimize total THz output by combining large-area illumination with thermal management strategies. The wavelength-independent nature of STEs provides inherent compatibility with Yb-technology, potentially enabling milliwatt-level THz average powers that could establish spintronic THz emitters as viable alternatives for applications calling for ultra broad bandwidths.

Last but not least, spintronic THz emitters with moderate to high average power would tremendously benefit from advances in sensitive ultra-broadband detection methods, which we discuss in Section V.a. Spintronic emitters themselves have also been explored for this purpose[204] albeit with moderate sensitivity.

## IV.e Other novel techniques to increase efficiency

In all methods currently used to generate broadband THz transients as described above, the NIR-to-THz conversion efficiencies remain very low. We detailed in the previous section how increased repetition rates and average powers usually result in additional conversion efficiency reduction with different physical origins, resulting in common reported values in the order of 0.1% in highly optimized cases. This calls for other enhancement strategies than simply increasing the driving laser power. This is particularly the case when considering extremely high repetition rates, where pulse energies at hand are moderate in spite of large average power. We detail below some strategies that have recently started to be explored to improve the overall efficiency in the goal of reaching high average power.

### IV.e.i Intra-cavity active/passive enhancement

A promising approach for increasing overall efficiency of THz sources is to make use of intracavity enhancement both in active or passive laser resonators. This concept has been successfully deployed for the generation of XUV radiation using high harmonic generation in gases, another spectral region with very low conversion efficiencies[205–207]. Despite great promise, this has been less explored for THz generation. However, a few recent results show great potential and thus we summarize them here. These results mostly focused on optical rectification, however many of the concepts could potentially be extended to other THz generation methods.

With regards to active resonators, mode-locked laser oscillators based on free-space geometries – mostly solid-state lasers with laser gain in bulk or thin-disk form- typically couple out only 1–10% of their intracavity power. Therefore, one to two orders of magnitude higher average and peak power circulate intracavity and are available to drive nonlinear conversion processes. This concept was attempted early on using a PC emitter in a mode-locked Ti:sapphire oscillator[208], but with very limited average power and presenting only very limited advantages compared to using the direct output of the laser. Using Yb mode-locked lasers a first result using a Kerr-lens mode-locked Yb:CALGO bulk oscillator, used a GaP crystal intracavity. Operating at 80 MHz with ~22 W of intracavity power and 105-fs pulses, the system generated up to 150 µW of THz power with a bandwidth extending to 5.5 THz. A second configuration using shorter, sub-50-fs pulses produced spectra reaching 7 THz, albeit at a moderate power of 35 µW, due to the thinner GaP crystal used[209]. Despite these promising results, GaP is limited by strong thermal lensing and nonlinear absorption, which hinder average power scaling, especially under high-repetition-rate and high-average-power conditions (see Section IV.a.i). Further scaling of this approach was presented using a thin plate of LN (with thickness smaller than the group-to-phase velocity mismatch length) inside a high-power thin-disk laser. LN offers significantly higher damage threshold, lower loss and lower thermal effects, allowing for much higher average intracavity power. In Ref.[210], a 50-µm-thick LN plate inside a Kerr-lens mode-locked Yb thin-disk laser stably operating with 264-W intracavity power and 115 fs pulse duration was used. An average THz power of 1.3 mW at a 44.8-MHz repetition rate was demonstrated. Multiple future improvements can be expected using this method as kW average powers with 50-fs durations have been achieved intracavity with this laser technology[207].

One drawback of this approach is the required specialized expertise in high-power mode-locked lasers to optimize and maintain stable mode-locking in the presence of a lossy nonlinear crystal. An alternative approach is to use low power commercial mode-locked lasers to seed passive enhancement cavities (ECs). ECs allow standard watt-level Yb lasers to achieve multi kilowatt-level intracavity powers with immense advances realized in last years[211,212]. Early attempts of using this technique were made with very limited success[213] seeding with Ti:sapphire lasers. Recently, this technique was revisited using Yb-seeding, achieving a 240-fold power enhancement to drive optical rectification in the same thin LN plates at a remarkably high 1.9 kW of intracavity average power. The output power generated was moderate with 0.65 mW of THz power at 93-MHz repetition rate[214], limited mainly by the long driving pulse duration. This represents an additional promising direction that could see further improvements in the future.

## IV.e.ii Metamaterials with local enhancement

Metamaterials and metasurfaces offer a promising path for THz generation by using engineered micro-structures to locally enhance nonlinear optical processes beyond what is achievable in bulk materials. These devices enable both efficient THz emission and on-chip functionality such as beam shaping, polarization control, and spatial phase modulation, all within ultra-compact footprints. A significant advantage of metasurfaces lies in their ability to confine THz fields to deeply subwavelength volumes, thereby enhancing local field strengths and boosting nonlinear interactions even in extremely thin films.

In 2011[215], a very first attempt on nanostructure-based THz generation was demonstrated utilizing silver-nanoparticle arrays, providing the first experimental evidence that photoelectrons ejected by plasmon-mediated multiphoton excitation could generate THz radiation through ponderomotive acceleration in the inhomogeneous plasmon field. Hale et al.[216] demonstrated THz generation from a nanostructured GaAs metasurface only 160-nm thick, using femtosecond 800-nm excitation. Despite its extreme thinness, the metasurface yielded THz signals comparable in strength to those from a bulk 500 µm GaAs crystal under similar conditions. The observed THz emission was attributed primarily to surface nonlinearities and shift currents, rather than traditional bulk optical rectification. In[217] this concept was extended to a binary-phase InAs metasurface, where tailored nanoresonators enabled both generation and control of the THz wavefront. Under the same 800-nm pump wavelength, the metasurface achieved a significantly stronger THz signal than a reference bulk ZnTe emitter. Importantly, the design suppressed photocurrent-based contributions and enabled polarization-sensitive THz emission with a controlled phase, marking a key advance in structured THz source engineering. Building on this, the same group developed an InAs metalens metasurface that simultaneously generated and focused THz pulses[218]. Regarding excitation wavelength compatibility, InAs—with its narrow bandgap of ~0.35 eV—is ideally suited for excitation at 1030 nm (photon energy ~1.2 eV), making these metasurfaces highly compatible with Yb-based laser systems.

By utilizing dielectric materials, the intrinsically high absorption of metal- and semiconductor-based metasurface THz emitters was reduced and a significant enhancement of the achieved efficiency was recently demonstrated[219]. Moreover, full control of the THz emission from a spintronic emitter was demonstrated by a spintronic metasurface[220], allowing for beam steering and full polarization control. Challenges and demands of various THz applications will potentially be addressed by further nanoengineering of these metasurfaces.

Overall, these metasurface THz emitters demonstrate strong potential for use in high-repetition-rate, Yb-laser-based systems due to their compact form factor, enhanced efficiency via local field effects, and integrated beam control functionality.

## V Challenges in detection schemes towards deploying high average power THz TDS in applications

High-average-power THz sources are in their infancy, and their potential for applications remains to be demonstrated. One of the most critical research areas for fully benefiting from these new, high-average power sources is the development of sensitive and fast broadband detection techniques tailored to the specific needs of a given application. We highlight here some areas of research that will support this.

To fully benefit from high-power Yb-based laser systems and thus high-power THz sources, detection must be improved or at least operate on similar noise floor as with low-power systems. The inherent disadvantage of high-power laser systems with their larger relative intensity noise (RIN) leads to larger amplitude fluctuations. Since the probe beam is derived from the main laser beam, these relative amplitude fluctuations can deteriorate the signal-to-noise ratio and thus the dynamic range. A model by Duvillaret et al. linked the RIN of the laser source to the SNR of a THz-TDS[221], but was developed for photoconductive emitter and receiver pairs. Generally, the literature reveals a significant gap in systematically characterizing the correlation between pump power scaling and final signal-to-noise mechanisms. This is likely because each specific application and its corresponding detection method require tailored optimization strategies, therefore will be explored in the specific context of their applications in the near future. Nevertheless, comprehensive studies evaluating mitigation strategies, such as passive noise reduction at the exact frequency required for a given detection scenario, piezo-based beam stabilization for pointing drift suppression, or noise-eater circuits for amplitude stabilization, represent an area of future work for these high-power lasers in this field.

Specific application areas, for example THz imaging, call for the fastest possible acquisition speeds. Emerging single-shot broadband THz detection techniques might significantly improve high-repetition-rate ultra-broadband THz sources[44]. Numerous approaches achieved single-shot detection by encoding the THz waveforms to the probe beams in space-to-time, angle-to-time, or frequency-to-time[46,221–224], and recently by parametric conversion of the THz waves[225]. The scalability towards data acquisition at kHz[45] and MHz[226] repetition rate has recently been demonstrated. However, additional work on improving single-shot SNR is required for these techniques to be on pair with scanning methods, which currently surpass them. Single-shot detections with high SNR are known to enable high-speed data acquisition rates in multidimensional pump-probe experiments[45], greatly enhancing studies in materials, chemical, or biological sciences, and all these applications could thus greatly benefit from their combination with high average power THz sources. Moreover, the recent demonstration of a THz photomultiplier tube as an instantaneous electric field detector for a Fourier-transform spectrometer has shown great potential for spectroscopic experiments without the requirement for optical gating[227].

On the other hand, the use of other methods to generate the time delay for TDS in architectures such as ASOPS[228], ECOPS[229], or SLAPCOPS[230] have also tremendously advanced throughout the last two decades, but so far have not been deployed with high average power drivers. In Ref.[231], an Yb dual-comb single-laser was deployed for THz-TDS in an ASOPS type configuration, but at low average power using standard photoconductive emitters and receivers. This is an area where also promising developments can be expected in the next few years by combining these concepts with high-power Yb technology.

The problem of fast and sensitive detection is significantly more severe for ultra-broad bandwidth sources >> 5 THz, for example organic crystals, spintronic THz emitters and plasma sources. Existing techniques that can capture the ultra-broad bandwidths usually exhibit low sensitivity. High repetition rates, as provided by Yb-driven systems, play therefore a particularly critical role in this area and sensitive detection methods are increasingly relevant to bring these sources to applications.

Most commonly, very thin crystals such as GaP or ZnTe are used for EOS to capture very broadband transients, but the small thickness limits achievable sensitivity and DR[190]. In this regard, organic crystals are promising, but their practical implementation for EOS has been limited by strong birefringence and poor crystal quality, leading to coherence loss during propagation through the detection crystal[136,232]. Several techniques have been demonstrated to circumvent this limitation by measuring the beam-profile modulation of the spatial probe beam profile[233], or probe pulse energy[234]. Another technique for broadband detection with organic crystals is the self-compensated birefringence by folded double propagation of the probe beam through the detection crystal[235]. However, this technique requires a quarter-wave plate that supports the full bandwidth of the probe beam, alongside high crystal bulk and surface quality, to minimize loss mechanisms such as scattering. Despite the significant progress in crystal quality by novel growth techniques, this limitation remains a subject for future improvements[136,232].

Recent demonstrations have established the potential of other high-quality organic crystals as broadband detectors even in the case of high average power detection[38]. In this study, the polarization of the probe and the THz fields were overlapped orthogonally in space and time in a 650-μm-thick MNA crystal for EOS detection. The authors demonstrated an all-organic-crystal THz-TDS with a bandwidth exceeding 8 THz with exceptional spectral density in this entire bandwidth. To evaluate the detection performance of the identical MNA crystal reported in Ref.[38], an ultra-broadband air-plasma source of the same group has also been used[40]. Figure 15 shows an example of EOS measurements of ~3600 traces acquired within 15 minutes, wherein the MNA crystal was able to coherently detect THz radiation with a high DR up to about 24 THz. The retrieved terahertz spectra reveal strong absorption at about 8.05 THz, 10.6 THz, and 19.3 THz. At about 1.5 THz, a peak DR of 62 dB has been reached, and in a mid-IR window between 14 THz and 19 THz, a DR of about 25 dB was achieved. The electrical field thus can be estimated to 186 kV/cm, which is significantly higher than the electric field estimated by the EOS measurements utilizing a 500-µm-thin GaP crystal[40]. This result demonstrates the potential of high-quality organic crystals for ultra-broadband applications.

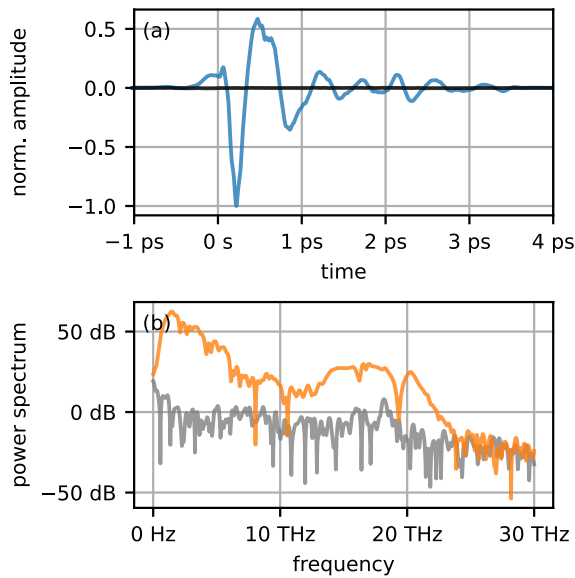

Figure 15: Ultra-broadband EOS detection of an air-plasma THz source based on a 650-µm thick MNA crystal. (a) time domain, (b) frequency domain.

For gapless ultra-broadband detection, the absorption dips in such crystals can represent a limitation depending on the application. In this regard, the most commonly used method is the air-biased coherent detection (ABCD) scheme that achieves coherent detection by introducing a local oscillator through four-wave mixing (FWM) of the THz probe, and an external electrical bias field[236] to isolate the THz electric field in the cross term of the FWM process, the bias electric field must be modulated at half the laser repetition rate, and detection is performed using a lock-in amplifier. The required bias field is on the order of several kV in air[149], presenting a significant technological challenge when operating at repetition rates beyond 100s of kHz. Furthermore, the ABCD scheme generally requires very high-energy probe beams, which become unfeasible at very high repetition rates. Several novel concepts have recently been introduced to address this challenge and will support future source development in this area. Fused silica's significantly higher third-order susceptibility and dielectric strength allowed for solid-state-biased coherent detection (SSBCD) at substantially lower probe pulse energies and bias voltages[237]. This reduction of the required modulated bias to 100s of V can be met by employing simple electronics, thus allowing for detection at significantly higher repetition rates. Alternatively, a novel balanced ABCD scheme has been introduced, leveraging the intrinsic advantage of balanced detection in analogy to EOS. This scheme achieves coherent detection by orienting the bias field orthogonally to the probe and THz fields[238]. This enables the acquisition of the field waveform at the full laser repetition rate, utilizing rapid scanning delay lines without bias modulation or lock-in detection. Moreover, this novel scheme outperformed conventional ABCD in terms of SNR and DR due to the noise reduction by the balanced detection. In another scheme, heterodyne detection is achieved by interfering with the THz field-induced second harmonic light from air with a local oscillator from a metal surface[239]. This method allows gapless detection up to 20 THz and can be performed at the full laser repetition rate, which promises to accelerate Yb-driven high repetition rate ultra-broadband THz sources.

## VI Metrology considerations

Finally, it is important to comment on the challenges of properly characterizing THz sources in general and how this impacts the case of high-power THz sources. Characterizing THz systems using average power as a single metric presents limitations. Multiple system parameters must be evaluated simultaneously to optimize experimental performance for a given application, in particular given the vast variety of bandwidths at hand from different setups. Another issue in the field that will support more source development is the standardization of power measurements: efforts in this direction have gained importance[240] as THz average power measurements achieve improved reliability[241–243]. The German metrology institute (PTB), for example, now provides calibration services at selected THz frequencies, enabling higher confidence in reported values and improved system comparisons when calibrated THz power meters are employed. We note also that as higher average become available and measurements operate further away from noise floors, spurious noise that were relevant when measuring average powers in the tens of µW, become less relevant.

Power spectral density represents another critical characterization parameter closely connected to THz average power, which can help in better comparing different sources. Recent standardization efforts have proposed methods for reporting relationships between recorded THz spectra and THz average power, enabling a first step toward absolute spectral calibration[244]. This standardization permits comparison between vastly different systems, such as narrowband lithium niobate sources versus broadband plasma-based generators. Systems exhibiting identical THz average powers can differ by orders of magnitude in power spectral density, highlighting distinct application domains.

Future standardization requires addressing two primary areas to enhance system comparability and interoperability. Current THz average power calibration capabilities are limited to a few selected narrowband THz frequencies, though these measurements can be linked to established calibration sources through traceable calibration chains. Expanding the calibration capabilities of metrological institutes across broader frequency ranges would significantly improve confidence in power spectral density measurements throughout the THz spectrum. Additionally, the community lacks consensus on standardized methods for reporting THz electric field strengths[18,245,35,246], where average power contributes alongside numerous additional parameters. Different calculation methodologies currently yield varying electric field values, creating challenges for experiments requiring precise field characterization. Achieving standardization levels already established for THz average power measurements demands continued community-wide coordination.

## Acknowledgments

The authors gratefully acknowledge Jeremy A. Johnson and David Michaelis from Brigham Young University for providing the experimental organic crystal MNA, and Mostafa Shalaby from Swiss Terahertz LLC for the organic crystal BNA used in our works. We thank Martina Havenith and Claudius Hoberg from Ruhr-Universität Bochum for valuable discussions, equipment sharing, and fruitful collaborative efforts that contributed to this work. We are grateful to Stephan Winnerl from Helmholtz-Zentrum Dresden-Rossendorf for providing the large-area emitters, and Tobias Kampfrath from Freie Universität Berlin for providing the spintronic THz emitters deployed in our measurements. We also thank Andreas Steiger from Physikalisch-Technische Bundesanstalt for insightful discussions regarding precise and accurate THz metrology techniques.

## Disclosures

The authors declare no conflict of interest.

## Data Availability Statement

Data underlying the results presented in this paper are available from the authors upon reasonable request.